\documentclass[aps,preprint,prd,nofootinbib]{revtex4-1}
\pdfoutput=1
\usepackage{graphicx}
\usepackage{psfrag}
\usepackage{float}
\usepackage{subfigure}
\usepackage{array}
\usepackage{graphicx}
\usepackage{amsmath}
\usepackage{amssymb}
\usepackage{mathrsfs}
\usepackage{bm}
\usepackage{ulem} 
\usepackage{slashed}
\usepackage{threeparttable}
\usepackage{multirow}
\usepackage[colorlinks, linkcolor=black, anchorcolor=black, citecolor=black]{hyperref}
\usepackage[amsmath,thmmarks,hyperref]{ntheorem}
\usepackage{soul}
\allowdisplaybreaks[4]

\begin{document}

\title{Spin-$1/2$ invisible particles in heavy meson decays}
\author{Geng Li$^1$\footnote{karlisle@hit.edu.cn}, Tianhong Wang$^1$\footnote{thwang@hit.edu.cn (Corresponding author)}, Yue Jiang$^1$\footnote{jiangure@hit.edu.cn}, Jing-Bo Zhang$^1$\footnote{jinux@hit.edu.cn} and Guo-Li Wang$^{1, 2, 3}$\footnote{gl\_wang@hit.edu.cn}\\}
\address{$^1$School of Physics, Harbin Institute of Technology, Harbin, 150001, China\\
$^2$Department of Physics, Hebei University, Baoding 071002, China\\
$^3$Hebei Key Laboratory of High-Precision Computation and Application of Quantum Field Theory, Baoding
071002, China}

\baselineskip=20pt

\begin{abstract}

The flavor-changing neutral current decay processes of the $B$ and $B_c$ mesons with the final states involving spin-$1/2$ particles are investigated. By considering the background of the Standard Model where $\nu\bar\nu$ contributes the missing energy and the experimental upper bounds for the branching fractions, we get the constraints of the coupling constants of the quark-antiquark and the assumed invisible particles $\chi\bar\chi$. The constraints of the coupling constants are then used to study the similar processes of the $B_c$ meson. At some specific region of $m_\chi$, the upper limit of BR($B_c\to D_{(s)}\chi\bar\chi$) is of the order of $10^{-6}$, while for BR($B_c\to D^\ast_{(s)}\chi\bar\chi$), it is $10^{-5}$. The possibility of distinguishing $\chi$ to be a Majorana or Dirac fermion by the differential branching fractions is also discussed.
	
\end{abstract}

\maketitle

\section{Introduction}

As the freeze-out mechanism~\cite{Bernstein:1985th, Srednicki:1988ce} can naturally interpret the observed dark matter abundance in our Universe, the weakly interacting massive particle (WIMP) is considered to be one of the most promising dark matter candidates. It is considered as a thermal relic from the local thermodynamic equilibrium early Universe~\cite{Izaguirre:2015yja}. The observed dark matter relic abundance $\Omega_c h^2 = 0.1131\pm 0.0034$~\cite{Bertone:2004pz, Komatsu:2008hk} sets a lower bound for WIMP's annihilation cross section. In specific models, the cross section can be connected to the mass of WIMP and coupling coefficients between WIMP and the Standard Model (SM) fermions. For example, the Lee-Winberg limit~\cite{PhysRevLett.39.165} demands its mass larger than a few GeV. However, this result is model dependent. With different models or proper selection of parameters, this constraint can be relaxed, which makes lower mass WIMP to be possible. For example, the MeV-scale light dark matter (LDM) was proposed~\cite{Pospelov:2007mp,Hooper:2008im} to explain the unexpected emission of 511 keV photons from the galaxy center. 

Previous experiments mostly focused on the dark matter particle with large mass, namely hundreds of GeV to several TeV. But recent experiment~\cite{Aprile:2017iyp} has set much stricter constraints on the parameter space for the WIMP with mass larger than several GeV. It provides a motivation to study the sub-GeV LDM through high-energy colliders. For example, CODEX-b at the LHCb experiment aimed to probe for GeV-scale long-lived particles~\cite{Gligorov:2017nwh}. Missing energy signals~\cite{Abdallah:2014hon} in flavor-changing neutral current (FCNC) processes of heavy mesons provide a possible way to probe light WIMP. Within the SM, neutrinos $\nu\bar\nu$ in the final state make contribution to the missing energy. However, theoretical calculations of the branching fractions of $B\to h_f\nu\bar\nu$ are less than the experimental bounds of $B\to h_f\slashed E$, where $h_f$ is the final meson and $\slashed E$ is the missing energy. So there is still some allowed parameter space for the decays involving other light invisible particles. 

Theoretically, spin of the invisible particle has several possibilities~\cite{Belyaev:2018pqr}. It can be a (pseudo)scalar~\cite{Boehm:2003hm}, a fermion~\cite{Kusenko:2009up}, or a hidden vector~\cite{Hambye:2008bq}. In the previous paper~\cite{Li:2018hgu}, we have considered the scalar and pseudoscalar cases. In this paper, we focus on the spin-$1/2$ light dark matter particles. There are many models involving the fermionic dark matter particles, such as sterile neutrino~\cite{Kusenko:2009up}, neutralino~\cite{Drees:2018dsj, Yue:2018hci}, Higgs-portal~\cite{Djouadi:2011aa, Djouadi:2012zc}, Z-portal~\cite{Arcadi:2014lta} and singlet-doublet~\cite{Hisano:2018bpz, Barman:2019aku, Restrepo:2019soi, Abe:2019wku, Fiaschi:2018rky}. Specifically, it can be either a Majorana or a Dirac fermion, as it is electrically neutral. Phenomenologically, the new invisible fermion can weakly interact with the SM fermions via a mediator, which can be a scalar~\cite{Matsumoto:2018acr}, pseudoscalar~\cite{Yang:2016wrl}, vector or axial-vector~\cite{Chala:2015ama} particle. The mass of the mediator is usually considered to be hundreds of GeV. In the energy level of heavy meson decays, namely several GeV, the branching ratios are greatly suppressed. However, as the FCNC and annihilation processes in the SM are also highly suppressed, the contribution of the new physics maybe important, which has been extensively studied in the decays of mesons~\cite{Bird:2004ts,Bird:2006jd,Badin:2010uh,Gninenko:2015mea,Barducci:2018rlx,Kamenik:2011vy,Bertuzzo:2017lwt}. For example, Ref.~\cite{Badin:2010uh} most focused on the $B$ meson annihilation processes, and Ref.~\cite{Kamenik:2011vy} studied various dark sectors in $B$ meson FCNC processes. 

In this work, we will further study the spin-1/2 invisible particles in $B$ and $B_c$ meson FCNC decays. Such studies for the $B_c$ meson are still missing. The $B_c$ meson has been massively produced and measured by the CDF~\cite{Aaltonen:2016dra}, ATLAS~\cite{Burdin:2016rzf}, CMS~\cite{Berezhnoy:2019yei}, and LHCb~\cite{Aaij:2019ths}  experiments. The production rate of $B_c$ meson on the LHCb Collaboration is close to 3.7 per mille of that of the $B$ mesons~\cite{Aaij:2019ths}. The $B_c$ events are in the order of $10^{10}$ per year. As the luminosity of the LHC increases significantly, much more $B_c$ events will be generated in the near future. We will first introduce the effective operators to describe the coupling between quarks and the invisible fermions.  The experimental upper bounds for the FCNC decay channels of the $B$ meson will then provide constraints of the coupling constants, which will be applied to calculate the upper bounds of the similar decay processes of the $B_c$ meson. To calculate the hadronic transition matrix elements, two methods are used: for the $B\rightarrow h_f$, the QCD light-cone sum rules (LCSR) is used, while for the $B_c\rightarrow h_f$, we apply the instantaneous Bethe-Salpeter (BS) method which is more suitable for such cases. For the light invisible fermions, both the Majorana and Dirac cases are considered. As they interact differently with quarks, for example, the Majorana fermion has neither vector nor tensor interactions, while Dirac fermion has both of them; the differential distribution will show slight difference. 
 
The paper is organized as follows: in Sec.~II, we present the model-independent effective Lagrangian to describe the coupling between the light invisible fermions and quarks, and extract the constraints of the coupling coefficients. In Sec.~III, we calculate the upper limits of the branching fractions of $B_c$ decays, and give the differential decay rate as a function of the missing energy. Finally, we draw the conclusion in Sec.~IV.

\section{Effective operators }
\begin{figure}[htb]
	\centering
	\includegraphics[width=1.1\textwidth,trim=50 545 0 140,clip]{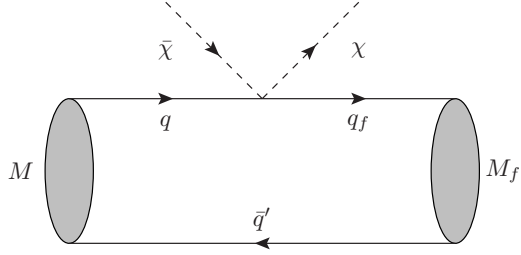}
	\caption{Feynman diagrams of decay channels involving invisible particles.}
	\label{Feyn01}
\end{figure}

The FCNC decay processes of heavy meson to spin-$1/2$ invisible particles $\bar\chi\chi$ are described in Fig.~\ref{Feyn01}, where $q$, $q_{_f}$, and $\bar q^\prime$ represent the quark and antiquark, respectively. 
The four-fermion vertex may be generated at the tree or loop level by introducing new physical mediators in specific models~\cite{Matsumoto:2018acr,Yang:2016wrl,Chala:2015ama}. In this work, we follow Ref.~\cite{Badin:2010uh} to introduce a model-independent effective Lagrangian,
\begin{equation}
\begin{aligned}
\mathcal L_{eff}=\sum_{i=1}^{9}{g_{_{fi}} Q_i},
\label{eq1}
\end{aligned}
\end{equation}
where the subscript $f$ represents fermion which can be Majorana or Dirac type, and $g_{fi}$ are the phenomenological coupling constants which are suppressed by the new physical energy scale $\Lambda$. There are nine independent effective operators $Q_i$s, which have the forms
\begin{equation}
\begin{aligned}
& Q_1=(\bar q_{_f}q)(\bar{\chi}\chi),~~~~~~~~~Q_2=(\bar q_{_f} \gamma^{5}q)(\bar\chi\chi),~~~~~~~~~ Q_3=(\bar q_{_f}q)(\bar{\chi}\gamma^{5}\chi),\\
& Q_4=(\bar q_{_f} \gamma^{5}q)(\bar\chi\gamma^{5}\chi),~~~Q_5=(\bar q_{_f} \gamma_{\mu}q)(\bar\chi\gamma^{\mu}\gamma^{5}\chi),~~~ Q_6=(\bar q_{_f} \gamma_{\mu}\gamma^{5} q)(\bar\chi\gamma^{\mu}\gamma^{5}\chi),\\
& Q_7=(\bar q_{_f} \gamma_{\mu}q)(\bar\chi\gamma^{\mu}\chi),~~~ Q_8=(\bar q_{_f} \gamma_{\mu}\gamma^{5}q)(\bar\chi\gamma^{\mu}\chi),~~~Q_9=(\bar q_{_f} \sigma_{\mu\nu} q)(\bar{\chi}\sigma^{\mu\nu} \chi).
\end{aligned}
\end{equation}

Two points should be stressed. First, in a specific model, a mediator maybe assumed to connect the SM quarks and the dark sector. In that case, not all the operators will contribute. Second, the coupling constants of some operators could be suppressed severely from a more fundamental point of view. For example, as Ref.~\cite{Kamenik:2011vy} pointed out, if one starts from an effective Lagrangian which respects the SM gauge symmetry, then the dimension-seven operators $(H^\dagger \bar q_{_f} q)(\bar \chi  \chi)$ and $(H^\dagger \bar q_{_f}\sigma_{\mu\nu} q)(\bar \chi \sigma^{\mu\nu} \chi)$ should be included. After electroweak symmetry breaking, they are reduced to $Q_1$ and $Q_9$, respectively, whose coefficients are suppressed by an additional factor $\frac{v}{\Lambda}$ with $v$ being the vacuum expectation value of Higgs field. 

 The upper limits of the coupling constants in the effective Lagrangian can be achieved by comparing the difference between theoretical predictions and the experimental data. As the corresponding detection of the $B_c$ meson is still missing, we cannot use the experimental data of $B_c$ meson to set constraints directly. Instead, the allowed region of the coupling constants can be obtained by considering the $B$ meson decay processes. These channels are $ B ^ {-} \to K ^ -(K^{\ast-}) + \slashed E $ and $ B ^ {-} \to \pi ^ -(\rho^-) + \slashed E $, which have the same vertex as that of the $B_c$ meson decays. The upper bounds of the $B$ meson decays involving missing energy are listed in the first column of Table~\ref{tab1}. These results are dependent on experimental accuracy. With more precise experimental conditions, these results might be further compressed in the future. Although we have cited the most stringent results so far, all of them leave room for contributions from new physics~\cite{Grygier:2017tzo}. The second column is the theoretical predictions, and the third one is the extracted upper limits for the decays involving the assumed particles. One notices that they are of the same order as that of the SM background. 
\begin{table}[htbp]
	\setlength{\tabcolsep}{0.5cm}
	\caption{The branching ratios (in units of $10^{-6}$) of $B$ decays involving missing energy.}
	\centering
	\begin{tabular*}{\textwidth}{@{}@{\extracolsep{\fill}}ccc}
		\hline\hline
		Experimental bound~\cite{Chen:2007zk,Grygier:2017tzo,Lai:2016uvj}&SM prediction~\cite{Kamenik:2009kc,Jeon:2006nq,Altmannshofer:2009ma,Bartsch:2009qp}&Invisible particles bound\\
		\hline
		${\rm BR}(B^\pm\to K^\pm\slashed E)<14$& ${\rm BR}(B^\pm\to K^\pm\nu \bar{\nu})=5.1 \pm 0.8$  & ${\rm BR}(B^\pm\to K^\pm\chi\chi)<9.7$	\\
		${\rm BR}(B^\pm\to \pi^\pm\slashed E)<14$& ${\rm BR}(B^\pm\to \pi^\pm\nu \bar{\nu})=9.7 \pm 2.1$  & ${\rm BR}(B^\pm\to \pi^\pm\chi\chi)<6.4$	\\
		${\rm BR}(B^\pm\to K^{*\pm}\slashed E)<61$& ${\rm BR}(B^\pm\to K^{*\pm}\nu \bar{\nu})=8.4 \pm 1.4$  & ${\rm BR}(B^\pm\to K^{*\pm}\chi\chi)<54$	\\
		${\rm BR}(B^\pm\to \rho^\pm\slashed E)<30$&${\rm BR}(B^\pm\to \rho^\pm \nu \bar{\nu})=0.49^{+0.61}_{-0.38}$  & ${\rm BR}(B^\pm\to \rho^\pm \chi\chi)<30$	\\
		\hline\hline
		\label{tab1}
	\end{tabular*}
\end{table}

\subsection{$\chi$ is a Majorana fermion}

We first consider the situation that the invisible particle is a Majorana fermion. In such a case, the vector and tenor currents give no contribution, namely, $\bar\chi\gamma^{\mu}\chi=0$ and $\bar\chi\sigma^{\mu\nu}\chi=0$ (these are not true for the Dirac fermion). For the $0^-\to0^-$ transitions, only three operators give nonzero contribution. The effective Lagrangian reads
\begin{equation}
\begin{aligned}
\mathcal L_{1}=g_{m1} (\bar q_{_f} q)(\bar\chi\chi)+g_{m3} (\bar q_{_f} q)(\bar\chi\gamma^{5}\chi)+g_{m5} (\bar q_{_f} \gamma_{\mu}q)(\bar\chi\gamma^{\mu}\gamma^{5}\chi),
\label{eq2}
\end{aligned}
\end{equation}
where the subscript $m$ in $g_{mi}$ indicates that we are dealing with Majorana fermions. The hadronic transition matrix elements can be expressed as,
\begin{equation}
\begin{aligned}
 \langle M_f^-|(\bar q_{_f} q) |M^-\rangle 
&= \frac{M^2-M_{f}^2}{m_q-m_{q_{_f}}}f_0 (s),\\
\langle M_f^-|(\bar q_{_f}\gamma_\mu q) |M^-\rangle 
&= (P+P_f)_{\mu}f_+ (s)+(P-P_f)_{\mu}\frac{M^2-M_{f}^2}{s} \big[f_0 (s)-f_+ (s)\big],\\
\langle M_f^-|(\bar q_{_f}\sigma_{\mu\nu} q) |M^-\rangle
&=i\big[P_{\mu} (P-P_f)_{\nu}-P_{\nu} (P-P_f)_{\mu}\big] \frac{2}{M+M_f} f_T(s),
\label{eq3}
\end{aligned}
\end{equation}
where $P$ and $P_f$ are the momenta of the initial or final mesons, respectively; $m_q$ and $m_{q_f}$ are the masses of quarks; $s$ is defined as $(P-P_f^{})^2$; $f_+$, $f_0$, and $f_T$ are form factors. Here we adopt the results of the LCSR method~\cite{Ball:2004ye} to write the form factors as, 
\begin{equation}
\begin{aligned}
&f_0(s)=\frac{r_2}{1-s/m_{fit}^2},\\
&f_{+,T}^K(s)=\frac{r_1}{1-s/m_R^2}+\frac{r_2}{(1-s/m_R^2)^2},\\
&f_{+,T}^{\pi}(s)=\frac{r_1}{1-s/m_R^2}+\frac{r_2}{1-s/m_{fit}^2},
\label{eq4}
\end{aligned}
\end{equation}
where the corresponding parameters $r_1$, $r_2$, $m_R$, and $m_{fit}$ are presented in Table \ref{tab2}.
\begin{table}[htb]
	\setlength{\tabcolsep}{0.5cm}
	\caption{Parameters in the form factors of the $B\to \pi(K)$ processes~\cite{Ball:2004ye}.}
	\centering
	\begin{tabular*}{\textwidth}{@{}@{\extracolsep{\fill}}ccccc}
		\hline\hline
		$F_i$&$r_1$&$r_2$&$m_{fit}^2$ (GeV$^2$)&$m_R$ (GeV)	\\
		\hline
		$f^K_0$&$0$&$0.330$&$37.46$&$\cdots$         \\
		$f^K_+$&$0.162$&$0.173$&$\cdots$&$5.41$    \\
		$f^K_T$&$0.161$&$0.198$&$\cdots$&$5.41$    \\
		\hline
		$f^{\pi}_0$&$0$&$0.258$&$33.81$   &$\cdots$    \\
		$f^{\pi}_+$&$0.744$&$-0.486$&$40.73$&$5.32$  \\
		$f^{\pi}_T$&$1.387$&$-1.134$&$32.22$&$5.32$  \\
		\hline\hline
		\label{tab2}
	\end{tabular*}
\end{table}

By finishing the three-body phase space integral, we get the branching ratio
\begin{equation}
\mathcal {BR}  = \frac{1}{512  \pi^3 M^3 \Omega\Gamma_{B^-}}\int\frac {ds}{s} \lambda^{1/2}(M^2, s, M_f^2)\lambda^{1/2}(s, m_\chi^2, m_\chi^2)\int d\cos\theta\sum_\lambda|\mathcal M|^2,
\label{eq5}
\end{equation}
where $\lambda(x, y, z)= x^2 + y^2 +z^2 -2xy-2xz -2yz$ is the K${\rm \ddot a}$llen function; $m_\chi$ is the mass of the invisible particle; $\theta$ is the angle between the three-dimensional momenta $\vec P_\chi$ and $\vec P_f$ in the center-of-momentum frame of the invisible particles; $\Gamma_{B^-}$ is the total width of $B^-$ meson; $\Omega=2$ originates from the final two invisible particles being identical (Majarana fermion), and $\Omega=1$ when $\chi$ is the Dirac fermion. In the square of the amplitude, there are interference terms which come from the contribution of two different operators. These terms are proved to be zero when the invisible particles are (pseudo)scalars~\cite{Li:2018hgu} . However, they are not all zero when $\chi$ is a fermion, which makes the calculations much more complicated. We will also calculate these terms, and actually for some of them, the contribution cannot be ignored. 

\begin{figure}[htbp]
	\centering
	\subfigure[~$B^- \to K^-\bar\chi\chi$]{
		\label{width-1}
		\includegraphics[width=0.45\textwidth]{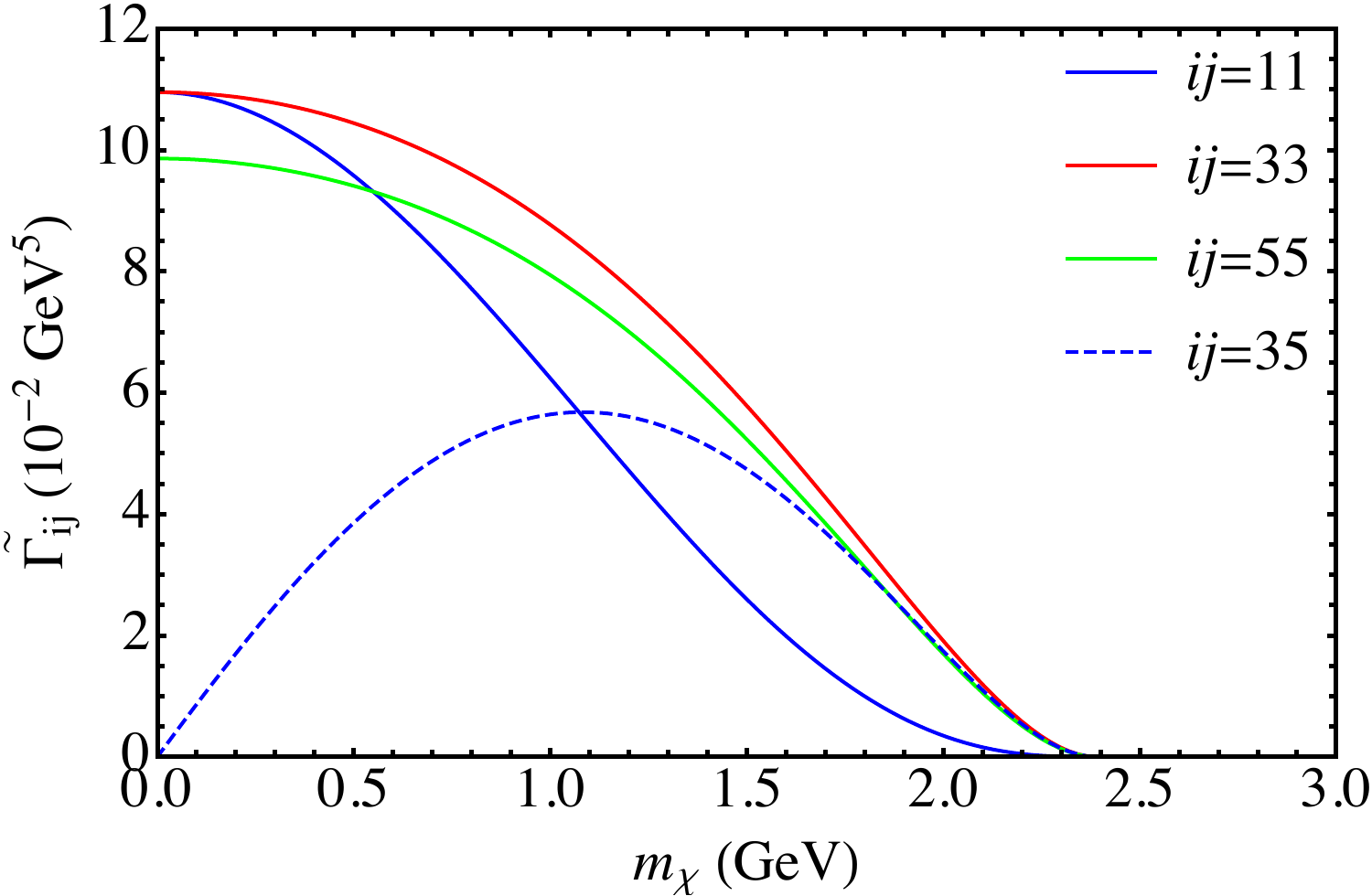}} 
	\hspace{0.5cm}
	\subfigure[~$B^- \to \pi^-\bar\chi\chi$]{
		\label{width-3}
		\includegraphics[width=0.45\textwidth]{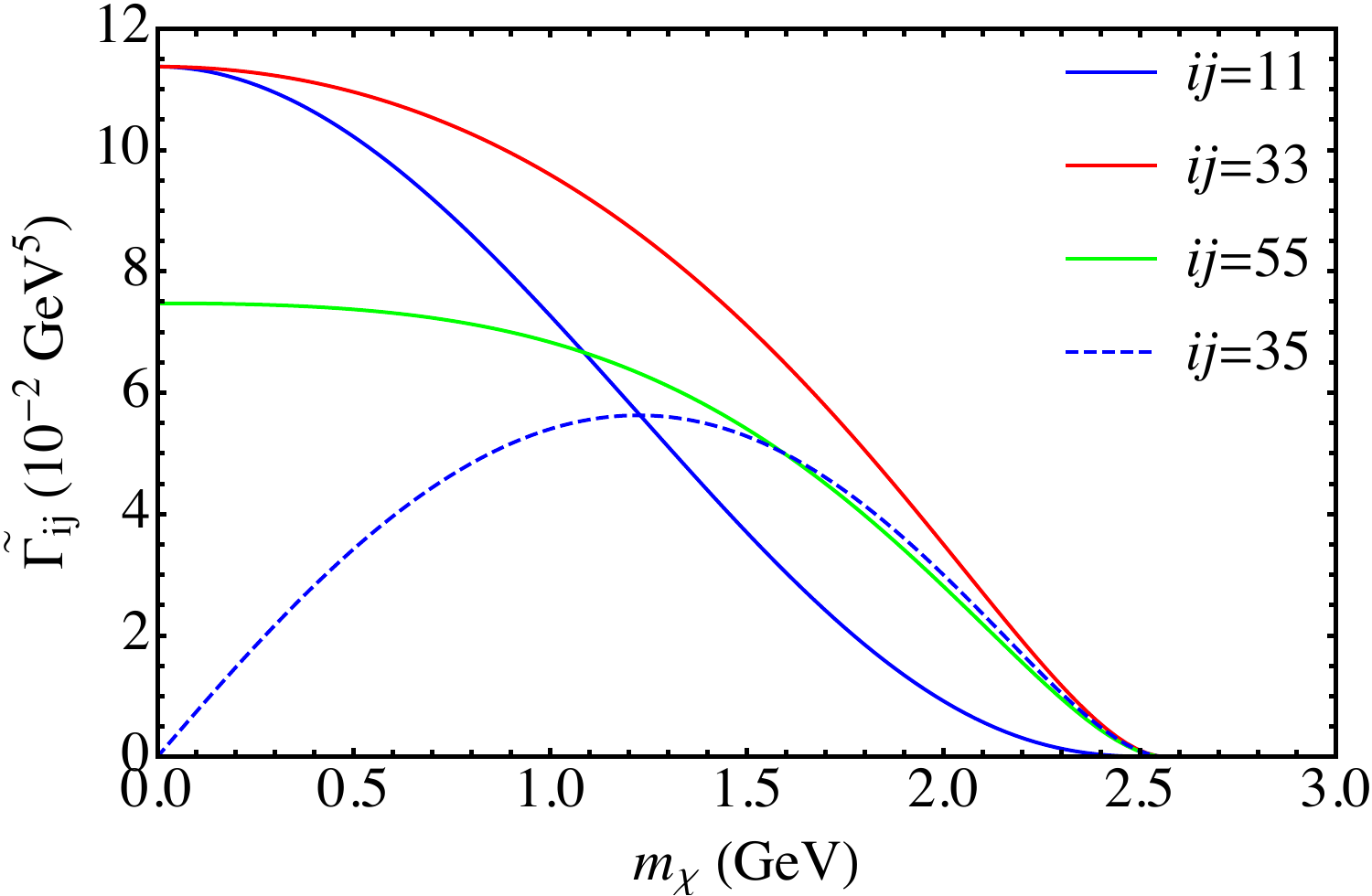}}
	\caption{$\tilde\Gamma_{ij}$ for $B\rightarrow K(\pi)\bar\chi\chi$ with $\chi$ being a Majorana fermion.}
	\label{width-1-3}
\end{figure}

The partial width can be written as
\begin{equation}
\begin{aligned}
\Gamma =\int {dPS_3 \big(\sum_{j} g_{mj}\mathcal{T}_j\big)^{\dagger} \big(\sum_{i} g^{ }_{mi} \mathcal{T}_i \big) }=\sum_{ij}g_{mj}g_{mi}\widetilde\Gamma_{ij},
\label{eq7}
\end{aligned}
\end{equation}
where we have taken $g_{ij}$ to be real for simplicity, and defined $\widetilde\Gamma_{ij}=\int dPS_3 \mathcal{T}^{\dagger}_j \mathcal{T}_i$, which is independent of the effective coupling constants. Some interference terms are zero by themselves or cancel each other out. The nonzero terms are $\widetilde\Gamma_{11}$, $\widetilde\Gamma_{33}$, $\widetilde\Gamma_{55}$, and $\widetilde\Gamma_{35}$. In Fig. \ref{width-1-3}, we plot them as functions of $m_\chi$. The solid and dashed lines represent the noninterference and interference terms, respectively. One can see that the two different channels have similar results, because the final mesons $K$ and $\pi$ have the same quantum number and small masses compared with that of the $B$ meson. The noninterference terms decrease when $m_\chi$ gets larger, because the phase space gets smaller. Detailed calculation shows that $\widetilde\Gamma_{11}$ and $\widetilde\Gamma_{33}$ are proportional to $(p_1\cdot p_2-m_{\chi}^2)$ and $(p_1\cdot p_2+m_{\chi}^2)$, respectively, where $p_1$ and $p_2$ are the momenta of two final invisible particles. So $\widetilde\Gamma_{11}$ is smaller than $\widetilde\Gamma_{33}$ except when $m_\chi=0$. $\widetilde\Gamma_{55}$ is less than $\widetilde\Gamma_{33}$ as they are related to different effective operators. The interference terms $\widetilde\Gamma_{35}$ and its complex conjugate $\widetilde\Gamma_{53}$ are numerically equal. One can see that they are zero when $m_\chi=0$ as they are proportional to $m_\chi^2$, which is quite different with $\Gamma_{ii}$. 

The effective Lagrangian for the $0^-\to1^-$  process has the form
\begin{equation}
\begin{aligned}
\mathcal L_{2}= g_{m2} (\bar q_{_f} \gamma^{5}q)(\bar\chi\chi)+g_{m4} (\bar q_{_f} \gamma^{5}q)(\bar\chi\gamma^{5}\chi)+g_{m5} (\bar q_{_f} \gamma_{\mu}q)(\bar\chi\gamma^{\mu}\gamma^{5}\chi)+g_{m6} (\bar q_{_f} \gamma_{\mu}\gamma^{5}q)(\bar\chi\gamma^{\mu}\gamma^{5}\chi).
\label{eq8}
\end{aligned}
\end{equation}
The hadronic transition matrix elements are parametrized by form factors $A_0$, $A_1$, $A_2$, $A_3$ $V$, $T_1$, $T_2$, and $T_3$~\cite{Straub:2015ica, Isgur:1990kf, Aliev:2010ki},
\begin{equation}
\begin{aligned}
 \langle M_f^{*-}|(\bar q_{_f}\gamma^5 q) |M^-\rangle 
&=-i\big[\epsilon \cdot (P-P_f)\big]\frac{2M_f}{m_q+m_{q_{_f}}}A_0(s),\\
\langle M_f^{*-}|(\bar q_{_f}\gamma_\mu\gamma^5 q) |M^-\rangle
&= i\bigg\{\epsilon_{\mu} (M+M_f)A_1(s)-(P+P_f)_{\mu}\frac{\epsilon \cdot (P-P_f)}{M+M_f}A_2(s)\\
&~~~-(P-P_f)_{\mu}  \big[\epsilon \cdot (P-P_f)\big] \frac{2M_f}{s} \big[A_3(s)-A_0(s)\big]\bigg\},\\
\langle M_f^{*-}|(\bar q_f \gamma_\mu q )|M^-\rangle
&=\varepsilon _{\mu \nu \rho \sigma} \epsilon ^\nu P^\rho (P-P_f)^\sigma\frac{2 }{M+M_f}V(s), \\
\langle M_f^{*-}|(\bar q_{_f}\sigma_{\mu\nu} q) |M^-\rangle,
&=i\bigg\{\varepsilon_{\mu \nu \rho \sigma} \epsilon^{\rho} (P+P_f)^{\sigma}T_1 (s) -\varepsilon_{\mu \nu \rho \sigma} \epsilon^{\rho} (P-P_f)^{\sigma}\frac{M^2-M_f^2}{s}\\
&~~~\times\big[T_1(s)-T_2(s)\big] -\varepsilon_{\mu \nu \rho \sigma} (P+P_f)^{\rho} (P+P_f)^{\sigma}\big[\epsilon \cdot (P-P_f)\big]\\
&~~~\times\big\{\frac{1}{M^2-M_f^2}T_3(s)-\frac{1}{s}\big[T_1(s)-T_2(s)\big]\big\}\bigg\},
\label{eq10}
\end{aligned}
\end{equation}
where $\epsilon$ is the polarization vector of the final meson, and the $\varepsilon^{0123}=1$ convention is used. 

The form factors are parametrized by~\cite{Straub:2015ica}, 
\begin{equation}
F_i(s) = P_i(s) \sum_k \alpha_k^i \,\left[z(s)-z(0)\right]^k,
\label{eq11}
\end{equation}
where the pole structure is $P_i(s)=(1-s/m_{R,i}^2)^{-1}$; $F_1$, $F_2$, $F_3$, $F_4$, $F_5$, $F_6$, $F_7$, and $F_8$ represent $A_0$, $A_1$, $A_{12}$, $A_3$, $V$, $T_1$, $T_2$, and $T_{23}$, respectively. $A_2$ and $T_3$ can be deduced from the relations
\begin{equation}
\begin{aligned}
A_3(s)&=\frac{M+M_f}{2M_f}A_1(s)-\frac{M-M_f}{2M_f}A_2(s)\\
A_{12}(s)&=  \frac{ (M+M_f){}^2 (M^2-M_f^2-s)A_1(s) - \big [(M+M_f)^2-s\big] \big[ (M-M_f)^2-s \big] A_2(s)  }{16 M M_f^2 (M+M_f)}\\
T_{23}(s)&=\frac{(M^2-M_f^2)(M^2+3M_f^2-s)T_2(s)-\big [(M+M_f)^2-s\big] \big[ (M-M_f)^2-s \big]T_3(s)}{8 M M_f^2(M-M_f)}. 
\label{eq12}
\end{aligned}
\end{equation}
$z(s)$ is defined as 
\begin{equation}
z(s) = \frac{\sqrt{s_+-s}-\sqrt{s_+-s_0}}{\sqrt{s_+-s}+\sqrt{s_+-s_0}},
\label{eq13}
\end{equation}
where
$s_\pm \equiv (M\pm M_f)^2$ and $s_0\equiv s_+(1-\sqrt{1-s_-/s_+})$. The related parameters are listed in Table \ref{tab3}.

\begin{table}[htb]
	\setlength{\tabcolsep}{0.5cm}
	\caption{Parameters in the form factors of the $B\to \rho(K^*)$ processes with $k_{\rm max}=2$~\cite{Straub:2015ica}.}
	\label{}
	\centering
	\begin{tabular*}{\textwidth}{@{}@{\extracolsep{\fill}}ccccc}
		\hline\hline
		$F_i$& $B\to K^*$ &$m_{R,i}^{b \to s}/$GeV& $B\to\rho$&$m_{R,i}^{b \to d}/$GeV \\
		\hline
		$\alpha_0^{A_0}$ & $0.36 \pm 0.05$ && $0.36 \pm 0.04$ &\\
		$\alpha_1^{A_0}$ & $-1.04 \pm 0.27$ &$5.366$& $-0.83 \pm 0.20$&$5.279$  \\
		$\alpha_2^{A_0}$ & $1.12 \pm 1.35$ && $1.33 \pm 1.05$ &\\
		\hline
		$\alpha_0^{A_1}$ & $0.27 \pm 0.03$ && $0.26 \pm 0.03$ &\\
		$\alpha_1^{A_1}$ & $0.30 \pm 0.19$ &$5.829$& $0.39 \pm 0.14$ &$5.724$\\
		$\alpha_2^{A_1}$ & $-0.11 \pm 0.48$ && $0.16 \pm 0.41$ &\\
		\hline
		$\alpha_0^{A_{12}}$ & $0.26 \pm 0.03$ && $0.30 \pm 0.03$ &\\
		$\alpha_1^{A_{12}}$ & $0.60 \pm 0.20$ &$5.829$& $0.76 \pm 0.20$  &$5.724$\\
		$\alpha_2^{A_{12}}$ & $0.12 \pm 0.84$ && $0.46 \pm 0.76$ &\\
		\hline
		$\alpha_0^{V}$ & $0.34 \pm 0.04$ && $0.33 \pm 0.03$ &\\
		$\alpha_1^{V}$ & $-1.05 \pm 0.24$ &$5.415$& $-0.86 \pm 0.18$ &$5.325$\\
		$\alpha_2^{V}$ & $2.37 \pm 1.39$ && $1.80 \pm 0.97$ &\\
		\hline
		$\alpha_0^{T_1}$ & $0.28 \pm 0.03$ && $0.27 \pm 0.03$ &\\
		$\alpha_1^{T_1}$ & $-0.89 \pm 0.19$ &$5.415$& $-0.74 \pm 0.14$ &$5.325$\\
		$\alpha_2^{T_1}$ & $1.95 \pm 1.10$ && $1.45 \pm 0.77$ &\\
		\hline
		$\alpha_0^{T_2}$ & $0.28 \pm 0.03$ && $0.27 \pm 0.03$ &\\
		$\alpha_1^{T_2}$ & $0.40 \pm 0.18$ &$5.829$& $0.47 \pm 0.13$ &$5.724$\\
		$\alpha_2^{T_2}$ & $0.36 \pm 0.51$ && $0.58 \pm 0.46$ &\\
		\hline
		$\alpha_0^{T_{23}}$ & $0.67 \pm 0.08$ && $0.75 \pm 0.08$ &\\
		$\alpha_1^{T_{23}}$ & $1.48 \pm 0.49$ &$5.829$& $1.90 \pm 0.43$ &$5.724$\\
		$\alpha_2^{T_{23}}$ & $1.92 \pm 1.96$ && $2.93 \pm 1.81$ &\\
		\hline\hline
		\label{tab3}
	\end{tabular*}
\end{table}

\begin{figure}[htbp]
	\centering
	\subfigure[~$B^- \to K^{*-}\bar\chi\chi$]{
		\label{width-5}
		\includegraphics[width=0.45\textwidth]{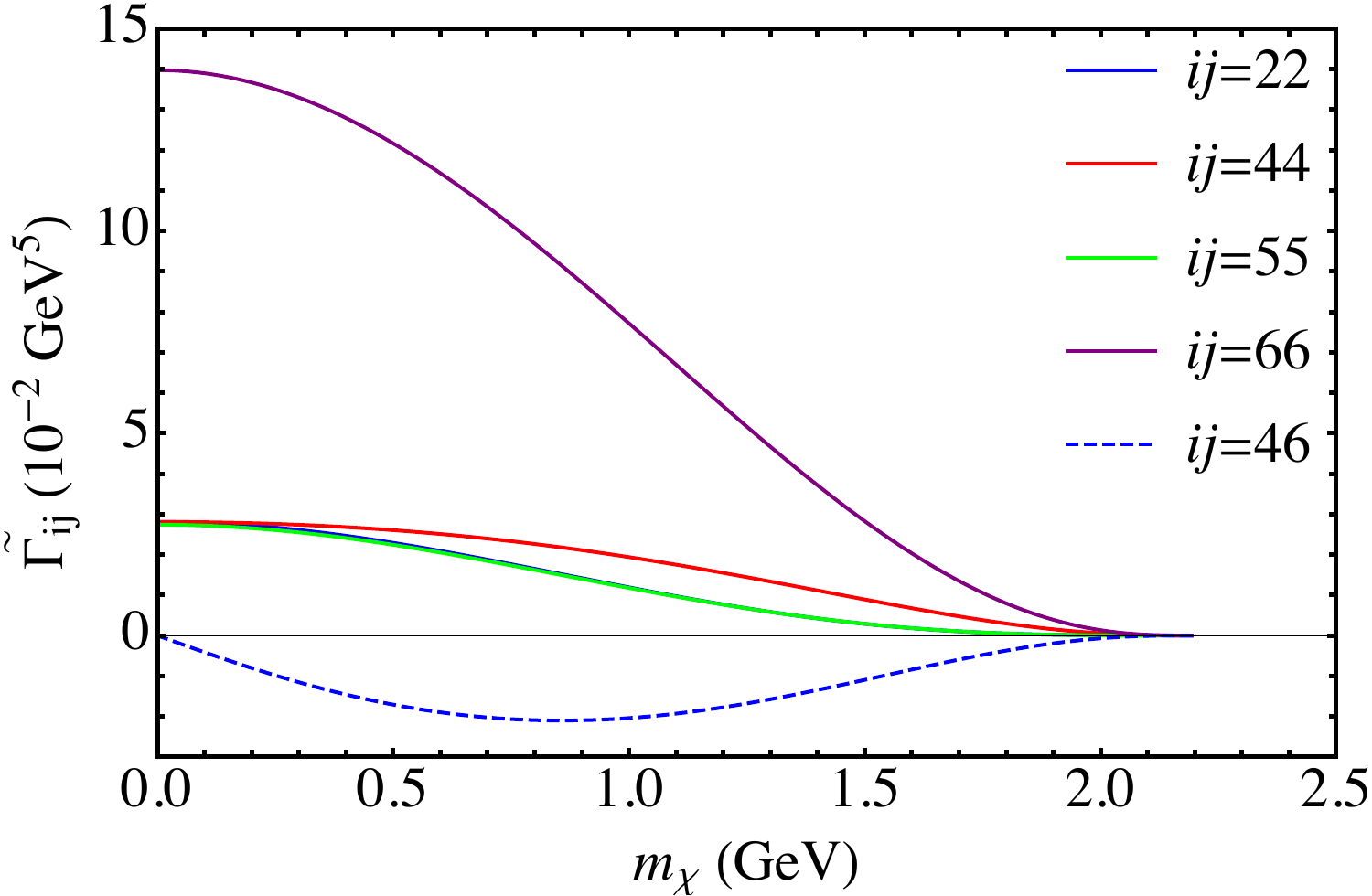}} 
		\hspace{0.5cm}
	\subfigure[~$B^- \to \rho^-\bar\chi\chi$]{
		\label{width-7}
		\includegraphics[width=0.45\textwidth]{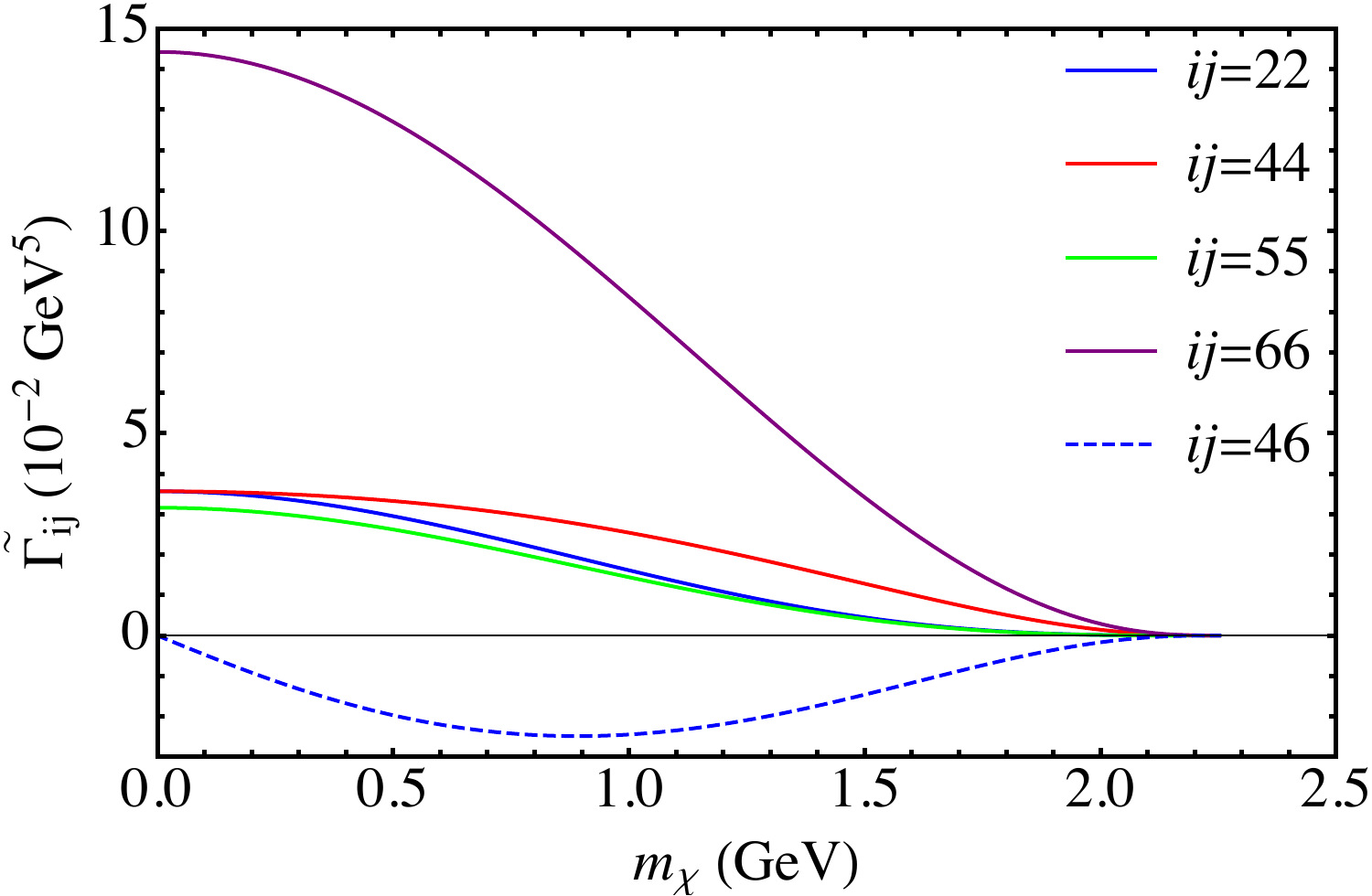}} 
	\caption{$\tilde\Gamma_{ij}$ for $B\rightarrow K^\ast(\rho)\bar\chi\chi$ with $\chi$ being a Majorana fermion.}
	\label{width-5-7}
\end{figure}

For $B\rightarrow K^\ast(\rho)\bar\chi\chi$, we plot $\widetilde\Gamma_{ij}$ as functions of $m_\chi$ in Fig.~\ref{width-5-7}. One notices that $\widetilde \Gamma_{66}$ is larger than the other terms. There is only one interference term $\widetilde\Gamma_{46}$ which is nonzero. Its contribution is negative. $\widetilde\Gamma_{22}$ and $\widetilde\Gamma_{55}$ are quite close to each other. For $B\rightarrow K^\ast\bar\chi\chi$, these two terms are almost coincident. 

\subsection{$\chi$ is a Dirac fermion}

\begin{figure}[htbp]
	\centering
	\subfigure[~$B^- \to K^-\bar\chi\chi$]{
		\label{width-2}
		\includegraphics[width=0.45\textwidth]{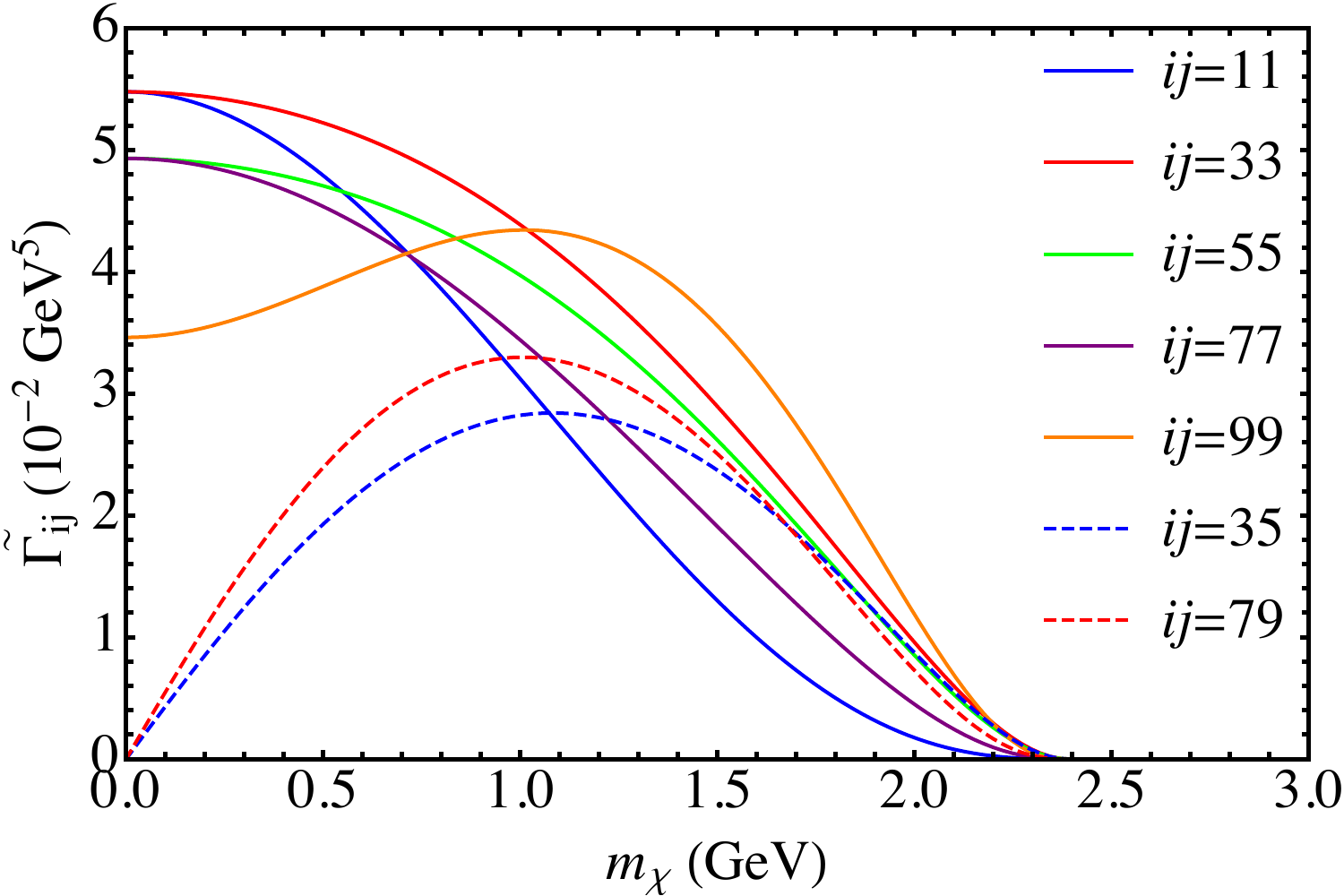}} 
	\hspace{0.5cm}
	\subfigure[~$B^- \to \pi^-\bar\chi\chi$]{
		\label{width-4}
		\includegraphics[width=0.45\textwidth]{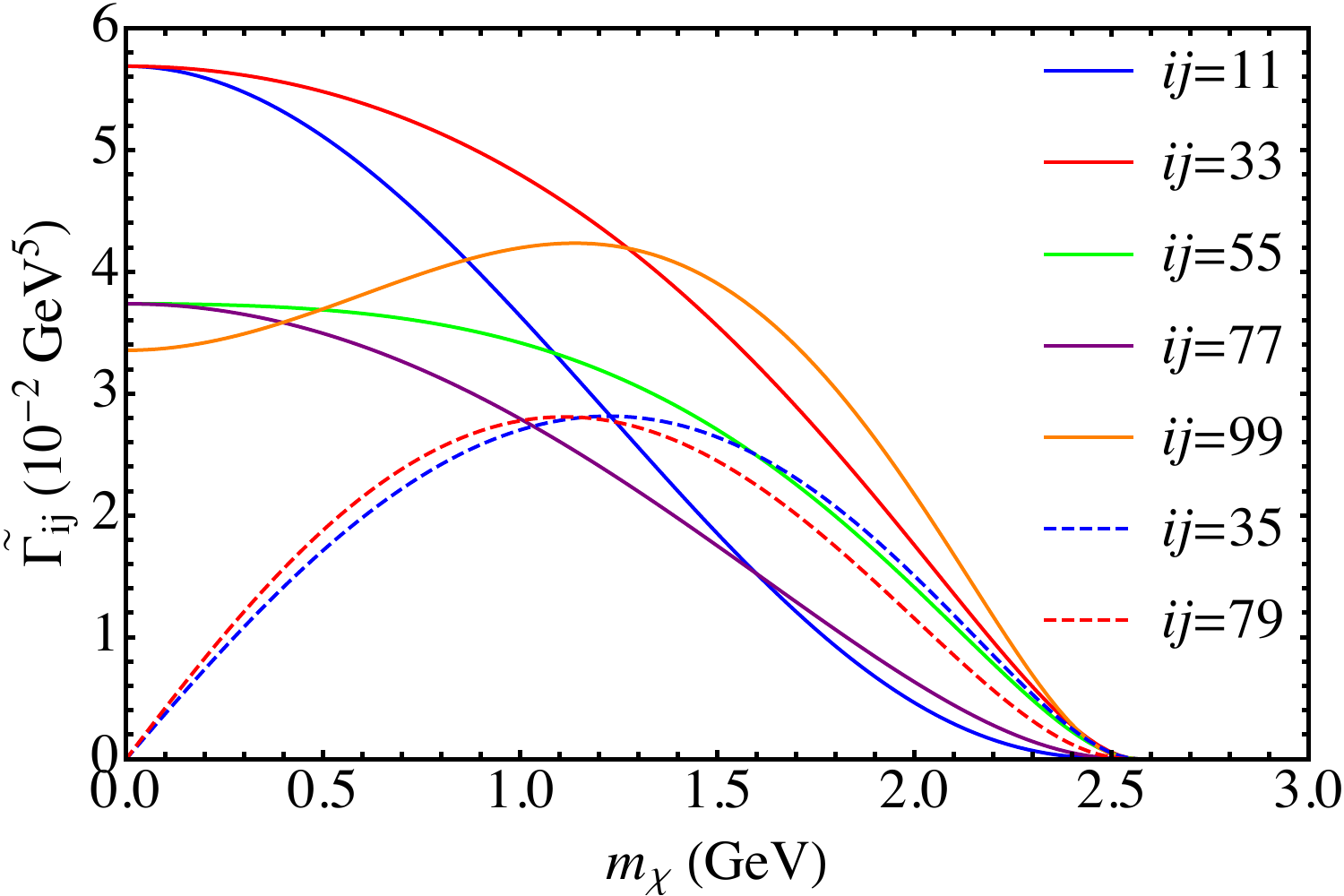}} 
	\caption{$\tilde\Gamma_{ij}$ for $B\rightarrow K(\pi)\bar\chi\chi$ with $\chi$ being a Dirac fermion.}
	\label{width-2-4}
\end{figure}

For the Majorana fermion, there is neither vector nor tensor interaction, while for the Dirac fermion, these two kinds of interactions also give contribution. When the invisible particle is a Dirac fermion, the effective Lagrangian has more operators. For the $0^-\to0^-$ transition, it can be written as
\begin{equation}
\begin{aligned}
\mathcal L_{3}=& g_{d1} (\bar q_{_f} q)(\bar\chi\chi)+g_{d3} (\bar q_{_f} q)(\bar\chi\gamma^{5}\chi)+g_{d5} (\bar q_{_f} \gamma_{\mu}q)(\bar\chi\gamma^{\mu}\gamma^{5}\chi)+g_{d7} (\bar q_{_f}\gamma_{\mu}q)(\bar\chi\gamma^{\mu}\chi)\\
&+g_{d9} (\bar q_{_f}\sigma_{\mu\nu}q)(\bar\chi\sigma^{\mu\nu}\chi),
\label{eq15}
\end{aligned}
\end{equation}
where $g_{di}$s are  the phenomenological coupling constants between the invisible Dirac fermions and quarks. $\widetilde\Gamma_{ij}$s are presented in Fig.~\ref{width-2-4}. One notices that they are about a half of that in the Majorana case where $\chi$ and its antiparticle $\bar\chi$ are identical. In Fig.~\ref{width-2-4}, one also notices that there are additional terms $\widetilde\Gamma_{77}$, $\widetilde\Gamma_{99}$ and $\widetilde\Gamma_{79}$, which represent vector and tensor currents, since they are not zero when $\chi$ is a Dirac fermion. Like above, $\widetilde\Gamma_{55}$ and $\widetilde\Gamma_{77}$ have same value when $m_{\chi}=0$. The $\widetilde\Gamma_{99}$ term increases first, then decreases to zero when the phase space gets less. The interference term $\widetilde\Gamma_{79}$ has the same trend as $\widetilde\Gamma_{25}$, for it is proportional to $m_{\chi}^2$.

\begin{figure}[htbp]
	\centering
	\subfigure[~$B^- \to K^{*-}\bar\chi\chi$]{
		\label{width-6}
		\includegraphics[width=0.45\textwidth]{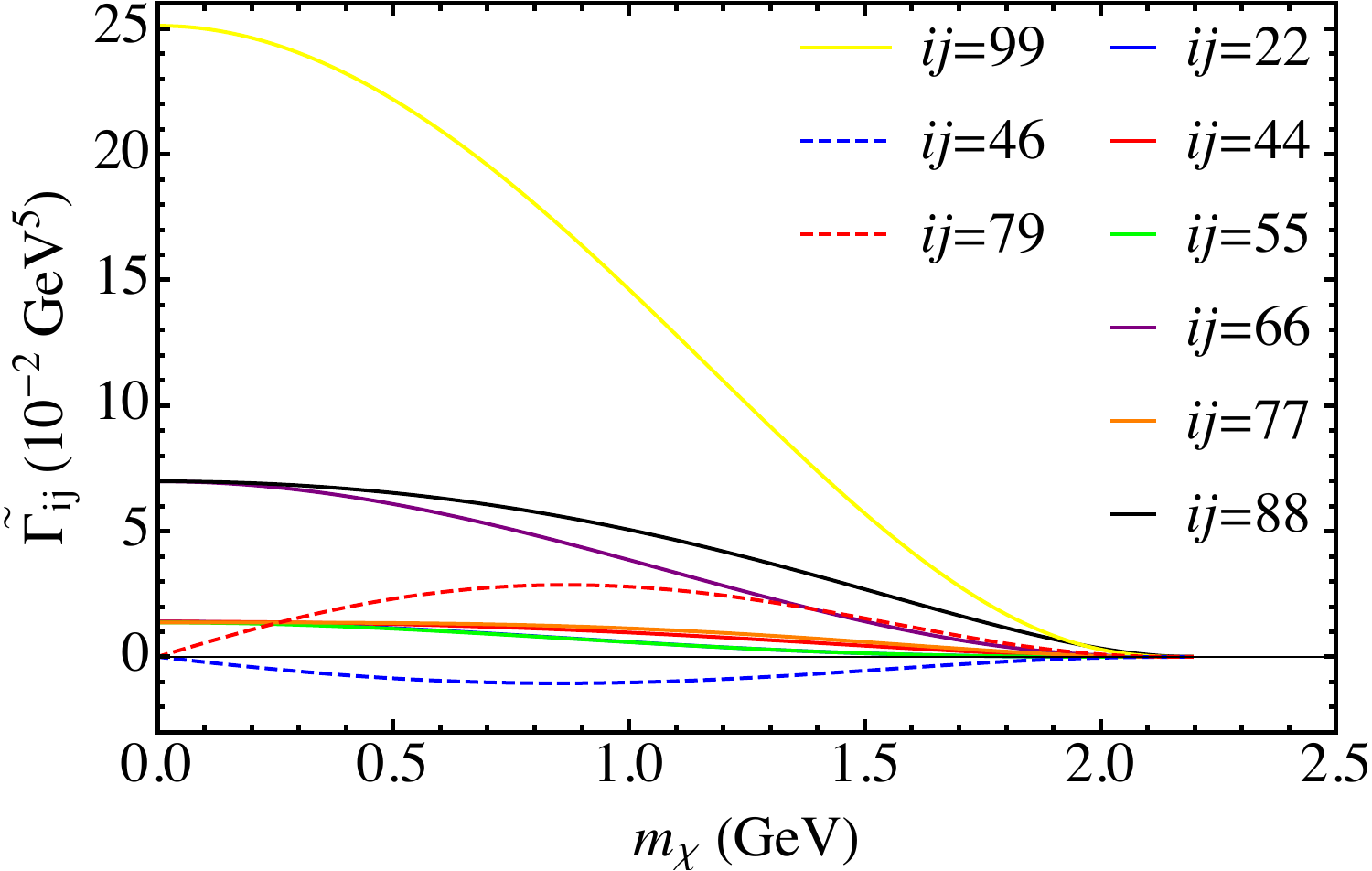}} 
	\hspace{0.5cm}
	\subfigure[~$B^- \to \rho^-\bar\chi\chi$]{
		\label{width-8}
		\includegraphics[width=0.45\textwidth]{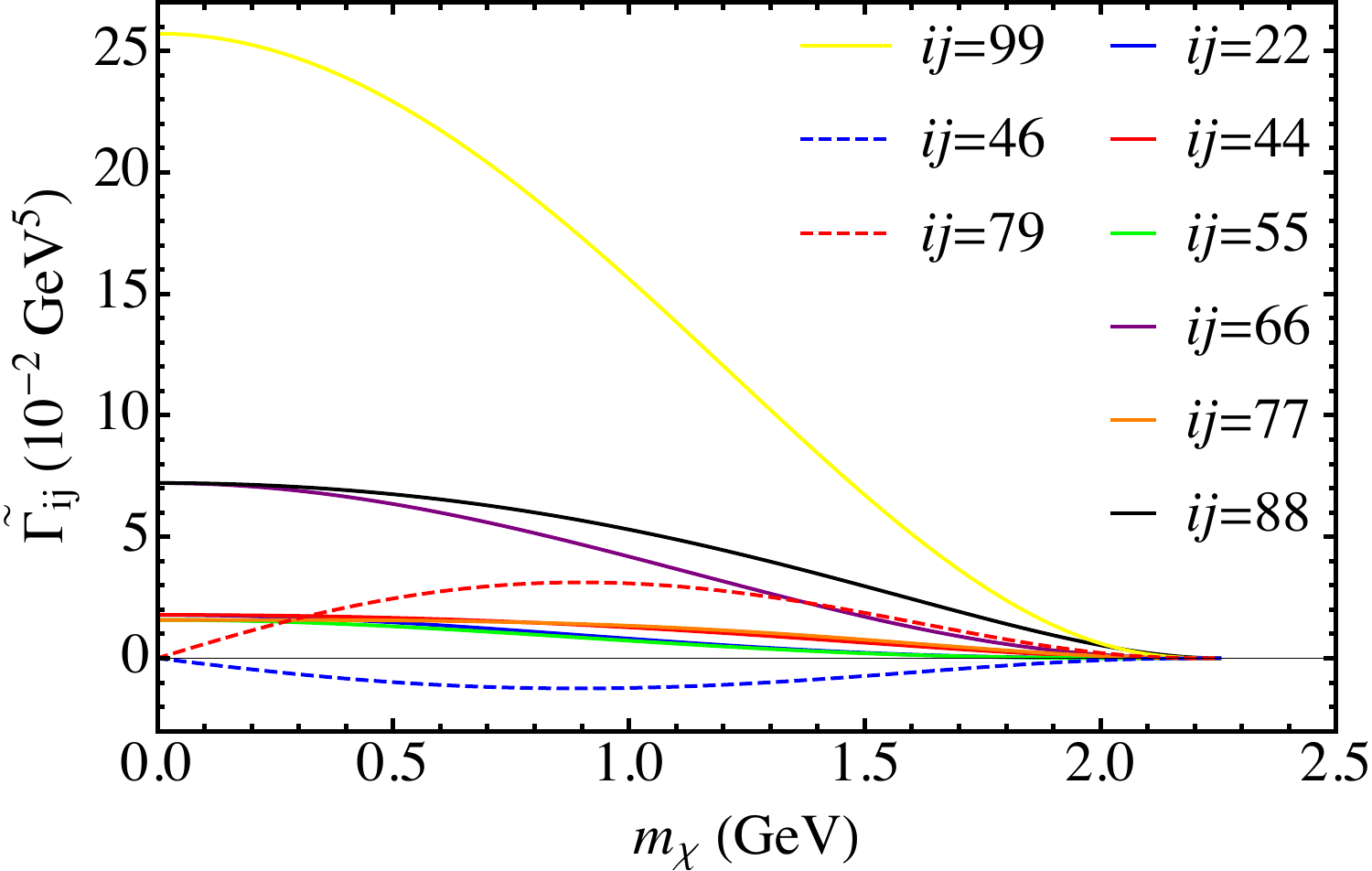}} 
	\caption{$\tilde\Gamma_{ij}$ for $B\rightarrow K^\ast(\rho)\bar\chi\chi$ with $\chi$ being a Dirac fermion.}
	\label{width-6-8}
\end{figure}

For $0^-\to1^-$ processes , the effective Lagrangian can be written as
\begin{equation}
\begin{aligned}
\mathcal L_{4}=&g_{d2} (\bar q_{_f} \gamma^{5}q)(\bar\chi\chi)+g_{d4} (\bar q_{_f} \gamma^{5}q)(\bar\chi\gamma^{5}\chi)+g_{d5} (\bar q_{_f} \gamma_{\mu}q)(\bar\chi\gamma^{\mu}\gamma^{5}\chi)+g_{d6} (\bar q_{_f} \gamma_{\mu}\gamma^{5}q)(\bar\chi\gamma^{\mu}\gamma^{5}\chi)\\
&+g_{d7} (\bar q_{_f}\gamma_{\mu}q)(\bar\chi\gamma^{\mu}\chi)+g_{d8} (\bar q_{_f}\gamma_{\mu}\gamma^5 q)(\bar\chi\gamma^{\mu}\chi)+g_{d9} (\bar q_{_f}\sigma_{\mu\nu}q)(\bar\chi\sigma^{\mu\nu}\chi).
\label{eq17}
\end{aligned}
\end{equation}
In Fig.~\ref{width-6-8}, we plot the nonzero $\widetilde\Gamma_{ij}$s as functions of $m_\chi$. One can see that the $\widetilde\Gamma_{ii}$ terms are not equal to zero even when $m_\chi$ takes zero. As before, the interference terms begin from zero and end up with zero when $m_{\chi}$ increases. Comparing with Fig.~\ref{width-5-7}, one can see that the additional term $\widetilde\Gamma_{99}$ related to $Q_9$ is larger than others. However, this does not mean that the tensor current makes much larger contribution to the partial width, for it also depends on $g_{f9}$, which might be suppressed compared with other coupling constants. 

\section{The decay modes of the $B_c$ meson}
In Sec.~II, the LCSR method is adopted to calculate the hadronic transition amplitude in the FCNC processes of $B$ meson, where the final meson is light. While for the $B_c$ meson decay modes, both initial and final mesons are heavy. Under these circumstances, the BS method is a good choice to calculate the hadronic transition amplitude. In this method, we can safely make an instantaneous approximation when solving the Bethe-Salpeter equation fulfilled by the wave functions of the heavy mesons. Details of how to solve the instantaneous BS equation can be found in~\cite{Kim:2003ny, Wang:2005qx}. The hadronic transition matrix element has the form
\begin{equation}
\begin{aligned}
\langle h^-|\bar q_1\Gamma^\xi b|B_c^-\rangle 
&= \int\frac{d^3 q}{(2\pi)^3} {\rm Tr}\left[\frac{\slashed P}{M}\overline\varphi_{P_f}^{++}(\vec q_{f})\Gamma^\xi\varphi_P^{++}(\vec q)\right],
\label{eq24}
\end{aligned}
\end{equation}
where $\varphi^{++}_P$ and $\varphi^{++}_{P_f}$ are the wave functions of the initial and final mesons, respectively; $\vec q$ and $\vec q_{_f}$ are the relative momentum of the quark and antiquark in the initial and final mesons, respectively. In the Standard Model, $\nu$ and $\bar\nu$ lead to the missing energy in the decay processes $B_c^-\to D_s^{(*)-}+\slashed E$ and $B_c^-\to B^{(*)-}+\slashed E$. The branching ratio of former channels is of the order of $10^{-7}\sim 10^{-6}$, while for the later ones, it is of the order of $10^{-15}\sim 10^{-14}$. The exact results can be found in our previous paper~\cite{Li:2018hgu}.

\subsection{$\chi$ is a Majorana fermion}

\begin{figure}[htbp]
	\centering
	\subfigure[~$B_c^- \to D_s^- \bar\chi\chi$]{
		\label{width-9}
		\includegraphics[width=0.44\textwidth]{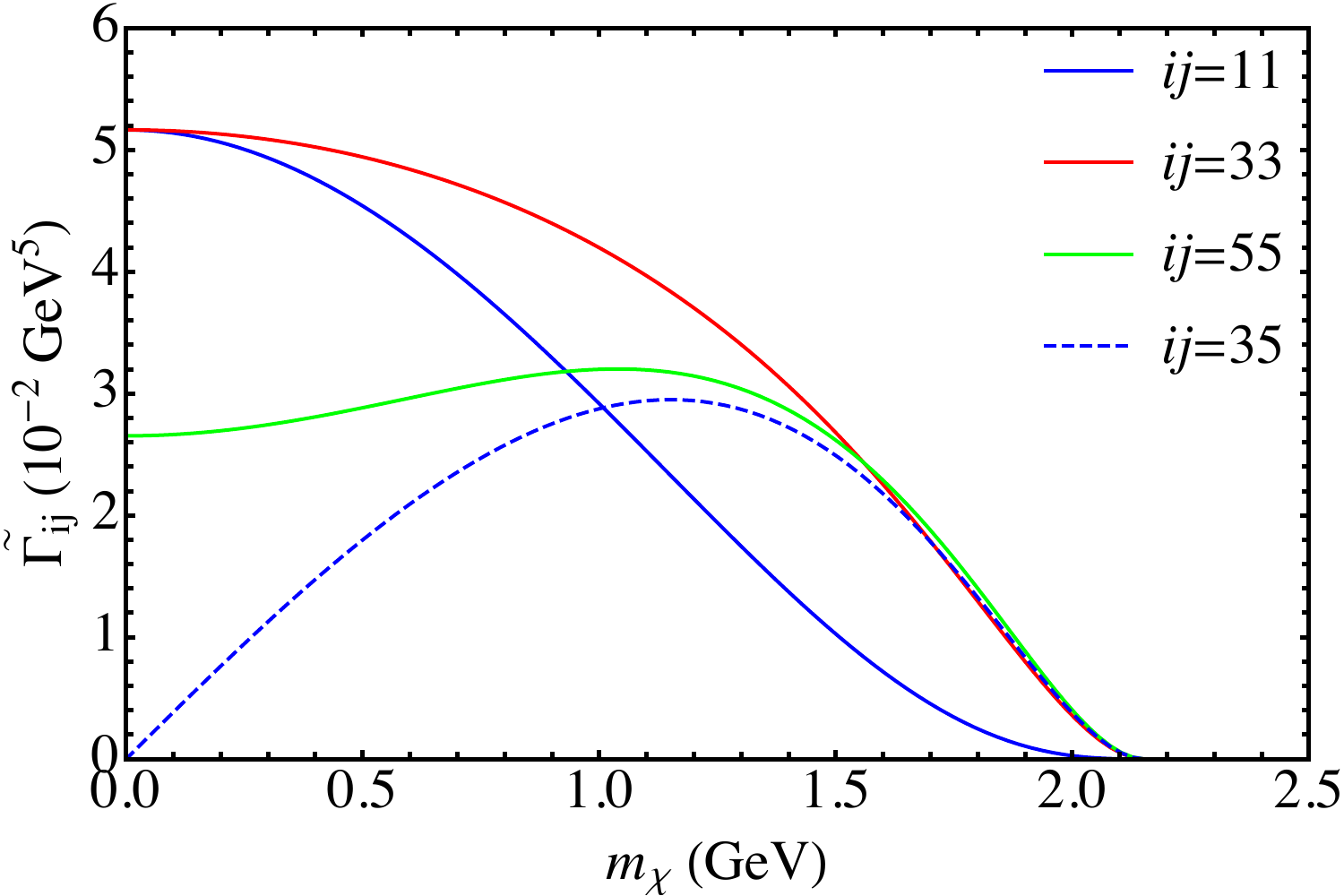}} 
	\hspace{0.5cm}
	\subfigure[~$B_c^- \to D_s^{*-} \bar\chi\chi$]{
		\label{width-15}
		\includegraphics[width=0.45\textwidth]{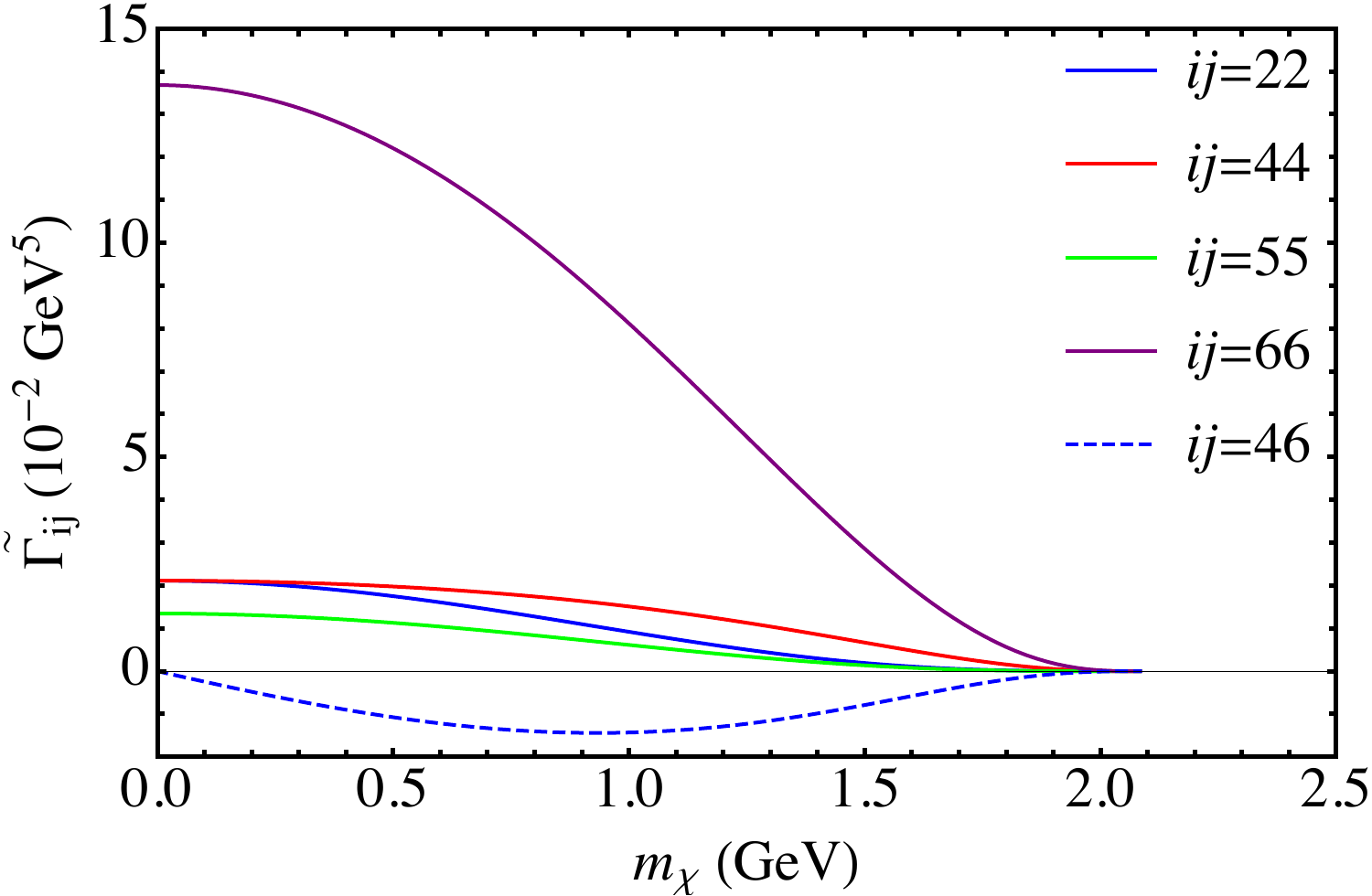}} \\
	\subfigure[~$B_c^- \to D^- \bar\chi\chi$]{
		\label{width-11}
		\includegraphics[width=0.44\textwidth]{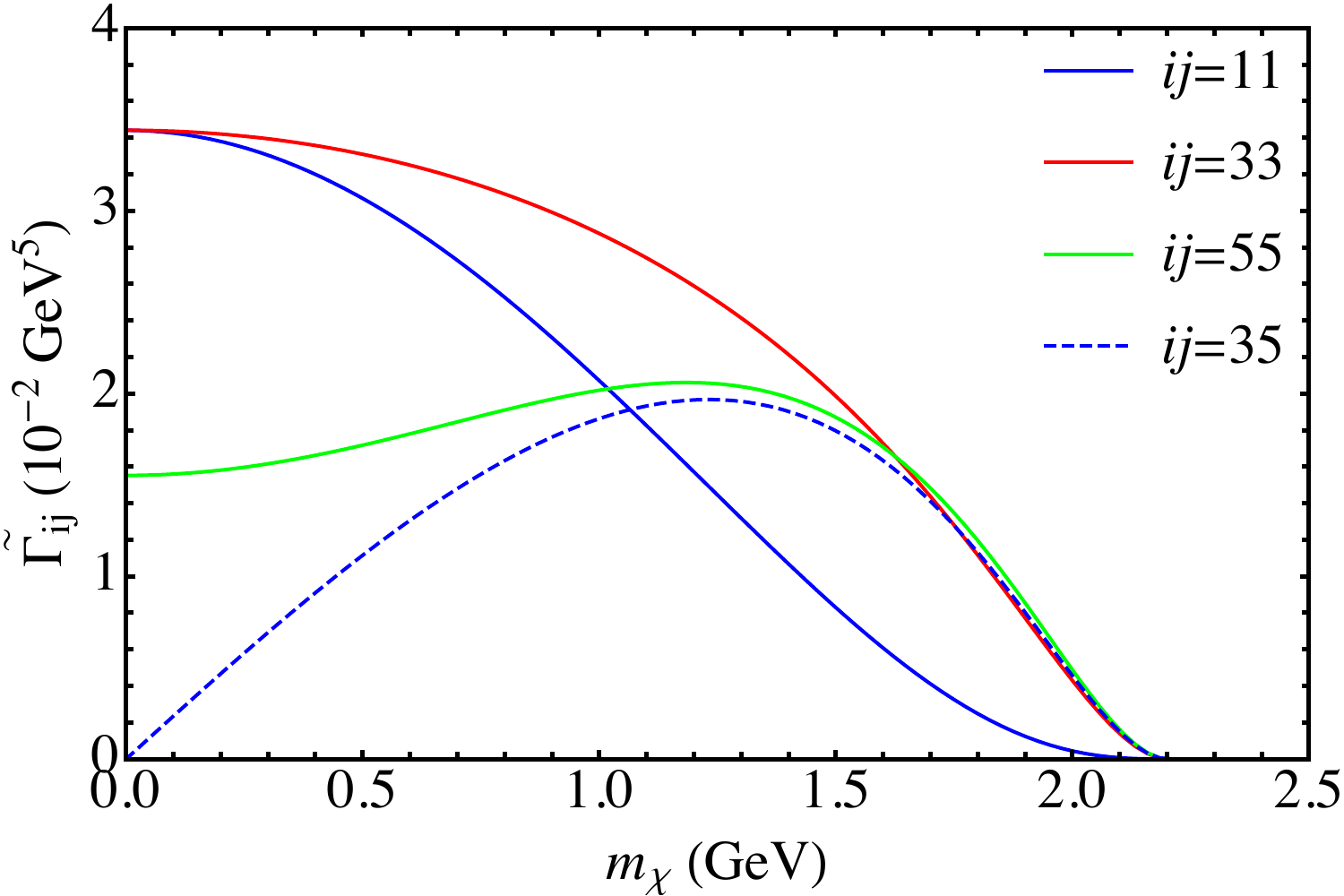}} 
	\hspace{0.5cm}
	\subfigure[~$B_c^- \to D^{*-} \bar\chi\chi$]{
		\label{width-17}
		\includegraphics[width=0.45\textwidth]{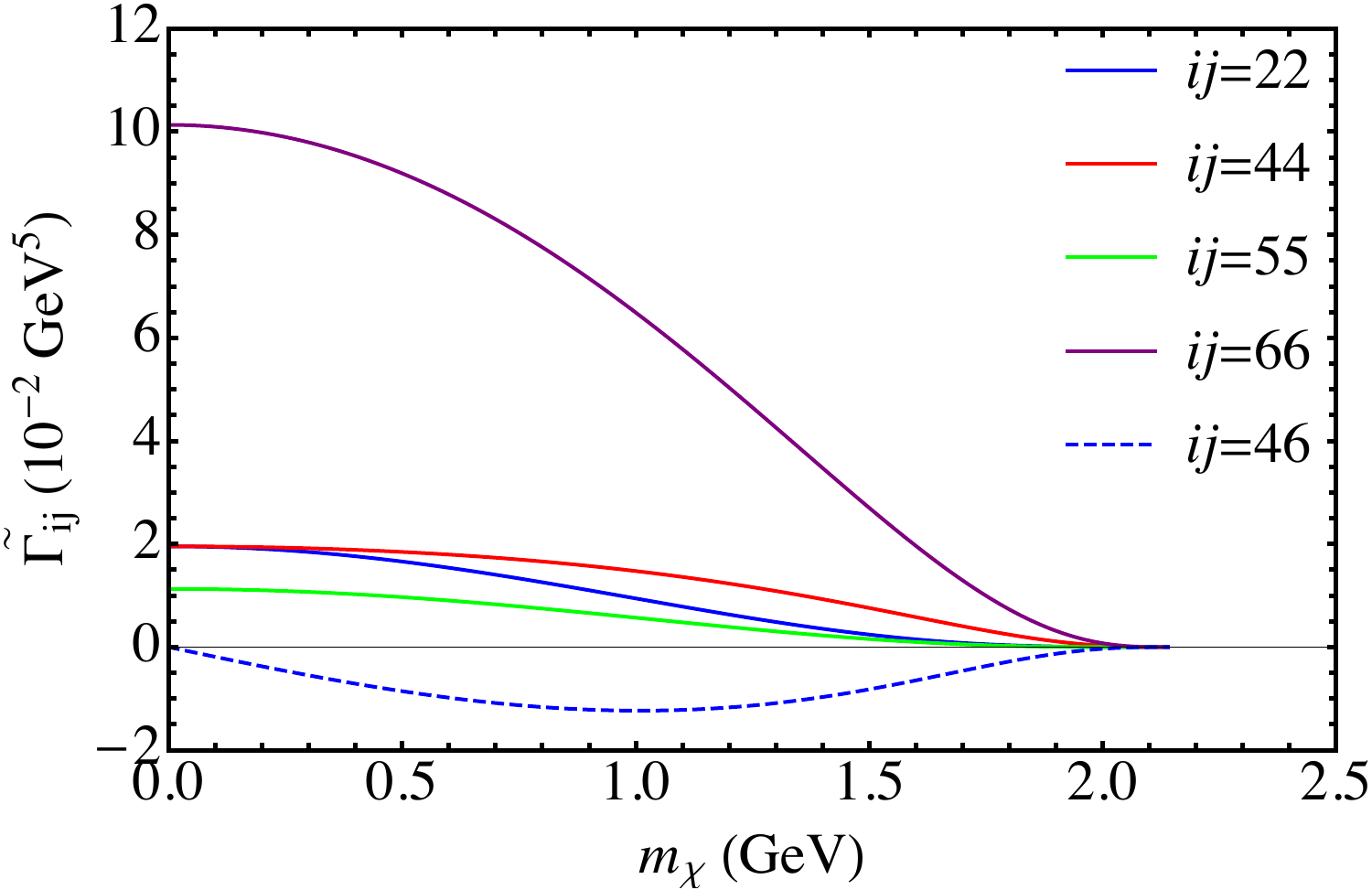}} \\
	\subfigure[~$B_c^- \to B^- \bar\chi\chi$]{
		\label{width-13}
		\includegraphics[width=0.44\textwidth]{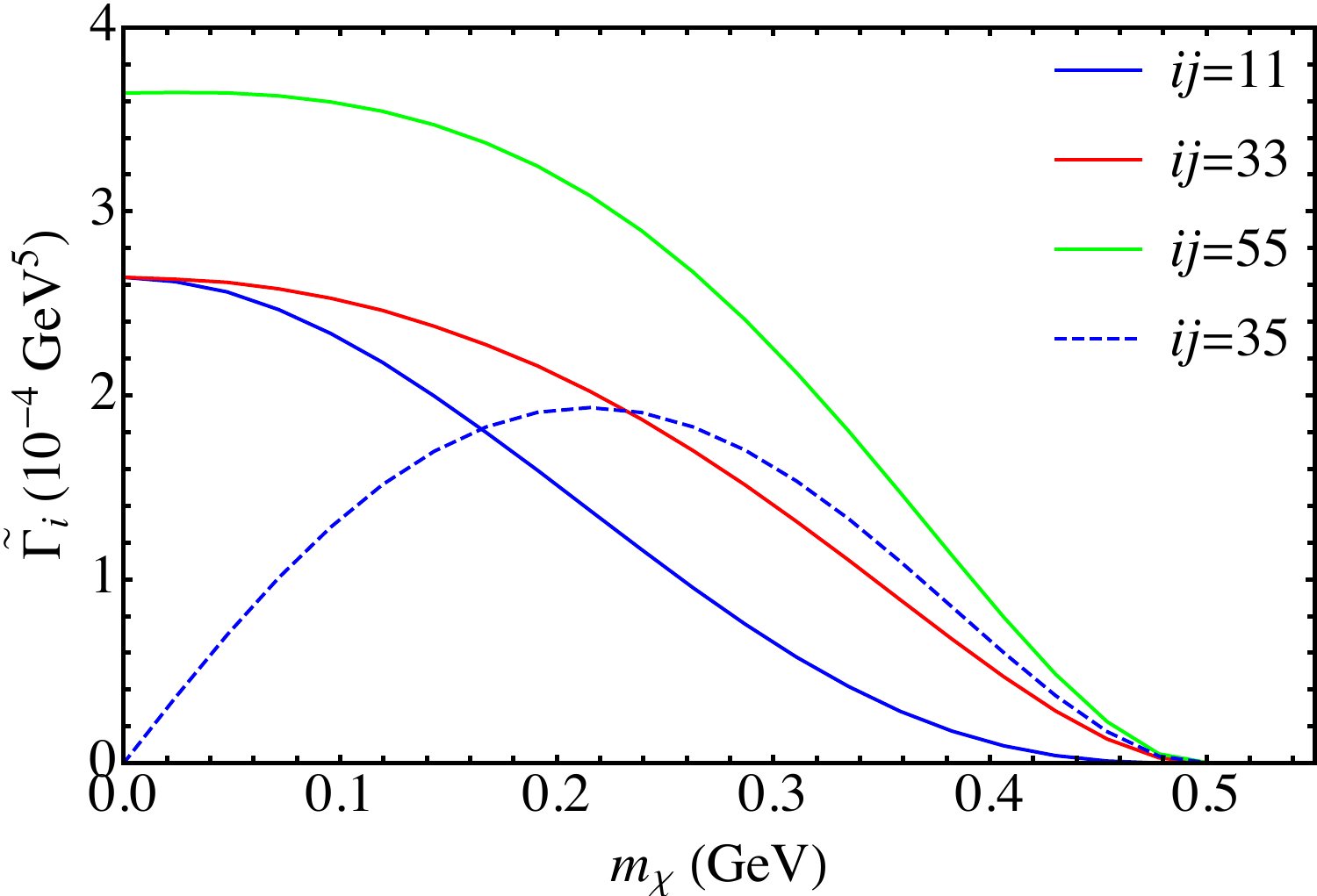}} 
	\hspace{0.5cm}
	\subfigure[~$B_c^- \to B^{*-} \bar\chi\chi$]{
		\label{width-19}
		\includegraphics[width=0.45\textwidth]{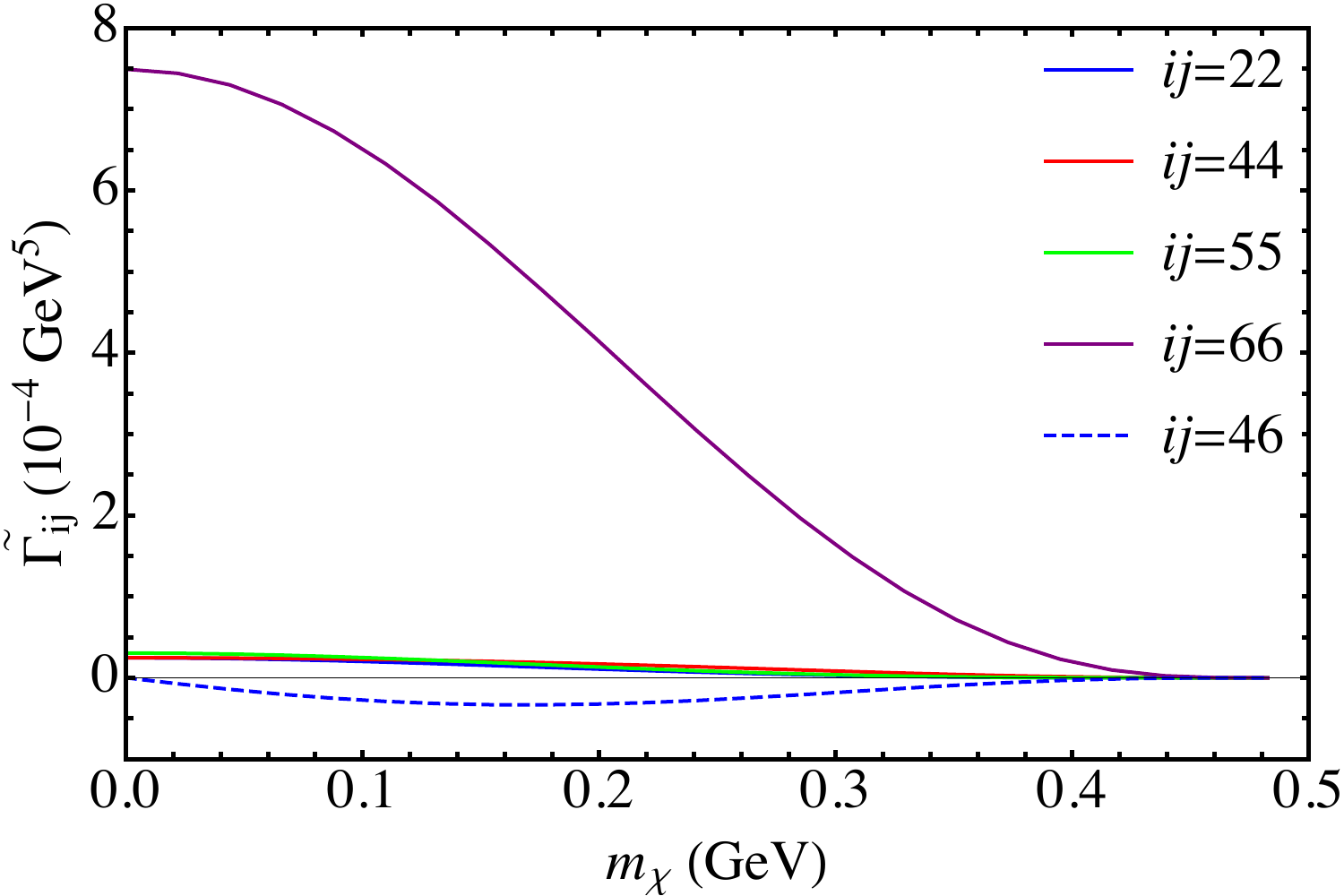}} \\
	\caption{$\tilde\Gamma_{ij}$ for $B_c\rightarrow h^{(\ast)}\bar\chi\chi$ with $\chi$ being a Majorana fermion.}
	\label{width-9-19}
\end{figure}

The decay processes of $B_c$ meson to Majorana fermions are also described by the effective Lagrangians in Eqs.~(\ref{eq2}) and (\ref{eq8}). Using Eq.~(\ref{eq24}), we get $\widetilde\Gamma_{ij}$s as functions of $m_\chi$, which are plotted in Fig.~\ref{width-9-19}. Although a different method is used to parametrize the form factors, the results of $B_c\rightarrow D_{(s)}$ and $B_c\rightarrow D^\ast_{(s)}$ are quite similar to those in Figs.~\ref{width-1-3} and \ref{width-5-7}, respectively, because the main difference of these channels comes from the different spectator quarks. We also consider the processes $B_c\rightarrow B^{(\ast)}\bar\chi\chi$. One notices that $\widetilde\Gamma_{ij}$s of the $c\to u$ processes are 2 orders of magnitude less than these of the $b\to d(s)$ processes. This is because the phase space of the former channel is less than that of the later one. The $\widetilde\Gamma_{55}$ term is smaller than $\widetilde\Gamma_{33}$ in $b$ decays while it is lager in $c$ decays, which means this operator is less sensitive to the phase space.

The next step is to set constraints for the coupling constants and calculate the upper limits for the branching fractions of $B_c$ decays. In Sec.II, we obtained $\widetilde\Gamma_{ij}$ for the decay processes of $B$ meson. Considering the upper limits of the branching fractions of such channels, we can extract the allowed parameter space for the effective coupling constants. Here we use two different ways to make the calculation. First, we assume just one effective coupling constant is nonzero, and its upper bound can be easily achieved. Of course, different operators will give different results. Second, we will scan the whole parameter space spanned by all the coupling constants under all the constraints. The first method sets maximum allowed region of each constant, namely, $g_{fi}$s. By second method we divide each regions into 400 bins. The program runs multiple bins and selects the maximum value of the branching ratio of $B_c$ meson when the selection of constants does not make the $\mathcal BR$ of $B$ meson beyond the experimental upper limit. 

\begin{figure}[htbp]
	\centering
	\subfigure[~$B_c^- \to D_s^- \bar\chi\chi$]{
		\label{br-1}
		\includegraphics[width=0.45\textwidth]{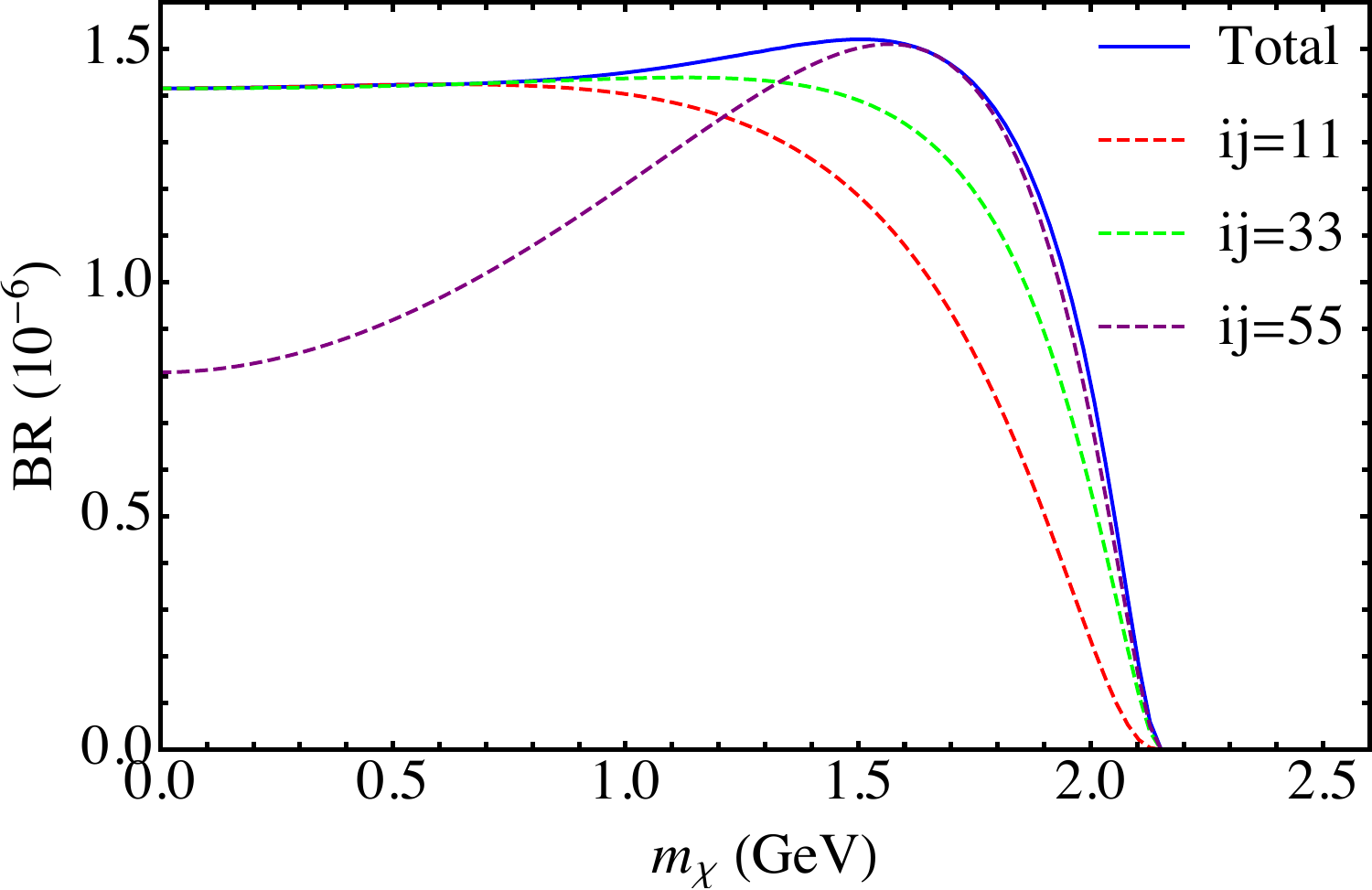}} 
	\hspace{0.5cm}
	\subfigure[~$B_c^- \to D^- \bar\chi\chi$]{
		\label{br-3}
		\includegraphics[width=0.45\textwidth]{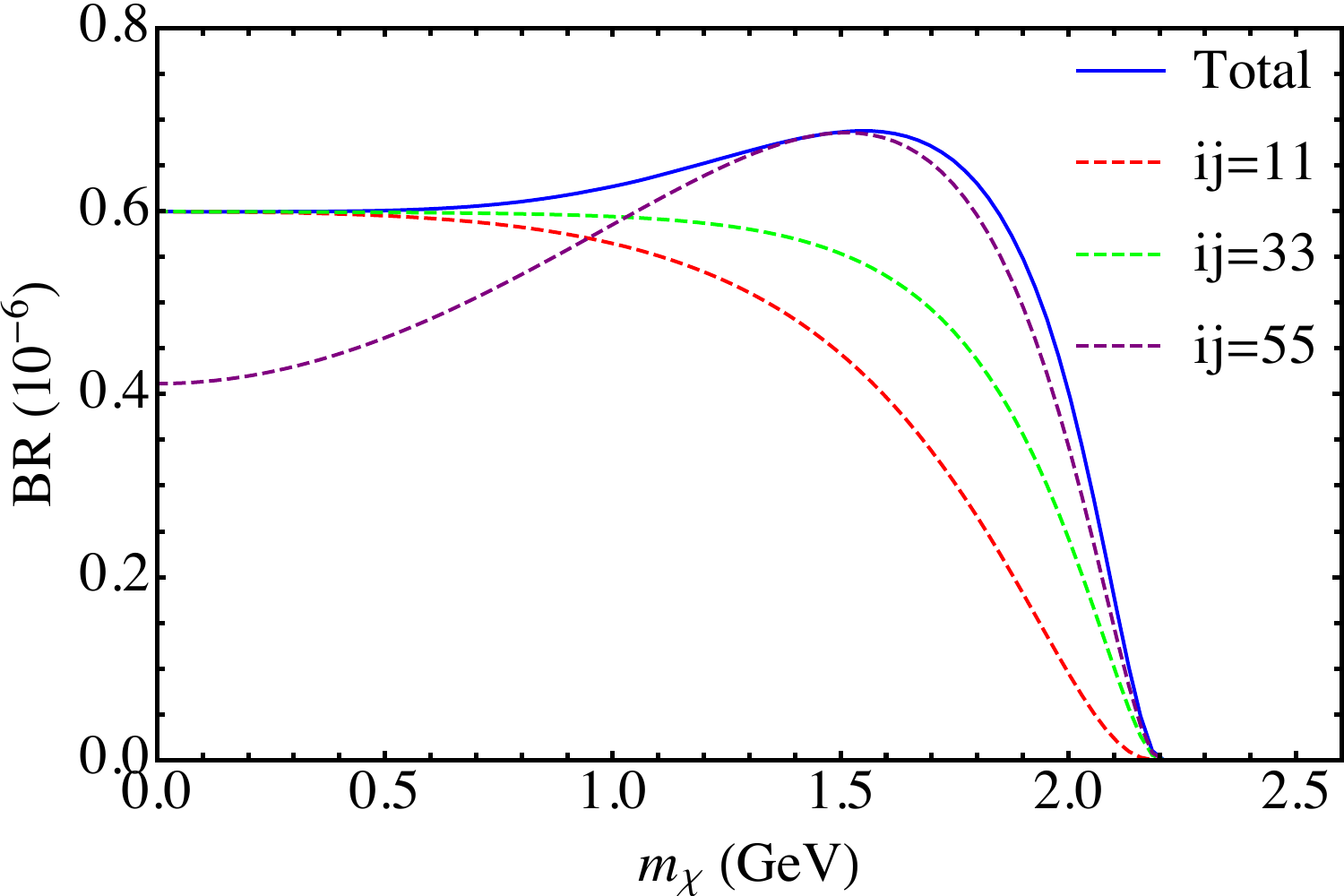}} \\
	\subfigure[~$B_c^- \to D_s^{*-} \bar\chi\chi$]{
		\label{br-5}
		\includegraphics[width=0.45\textwidth]{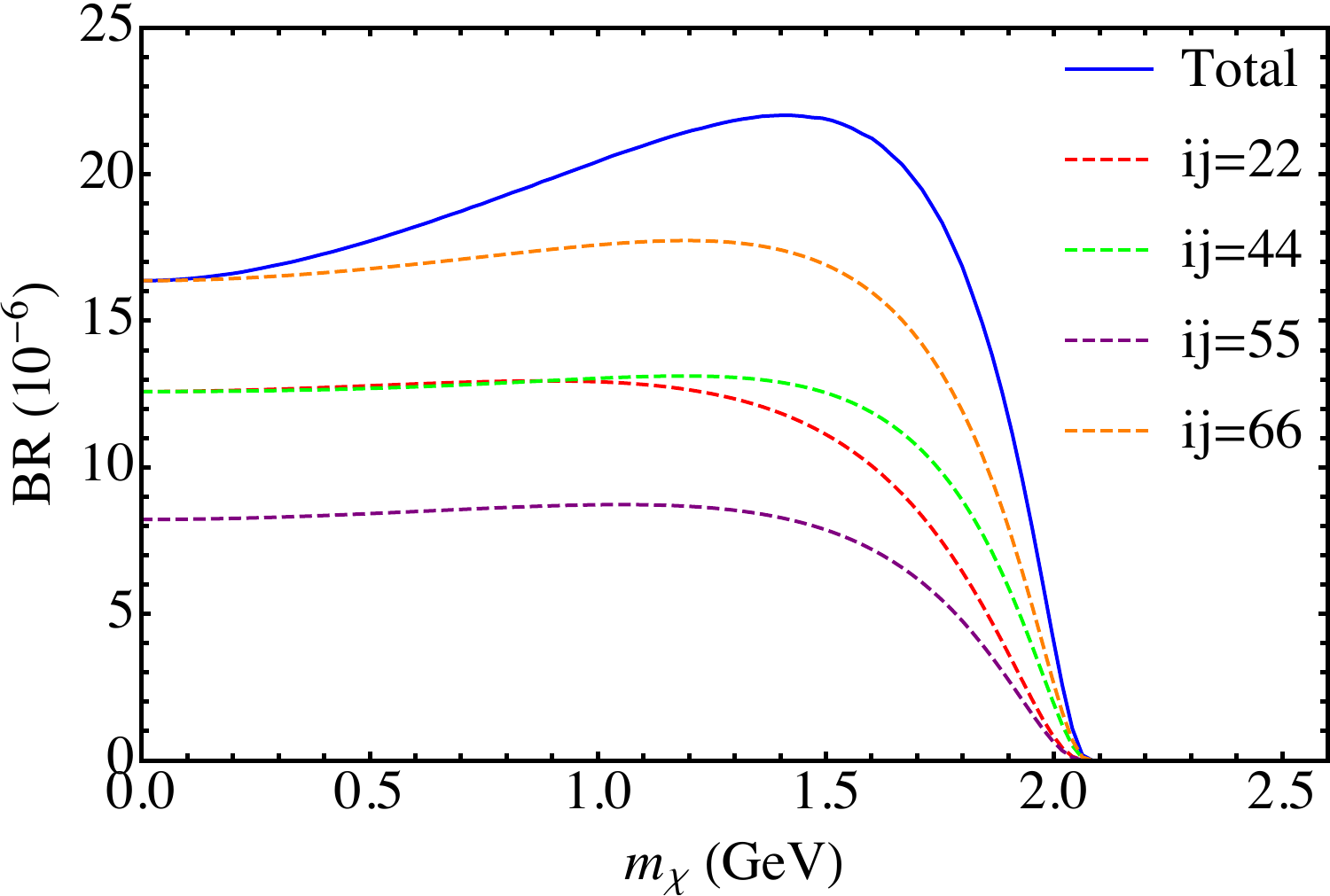}} 
	\hspace{0.5cm}
	\subfigure[~$B_c^- \to D^{*-} \bar\chi\chi$]{
		\label{br-7}
		\includegraphics[width=0.45\textwidth]{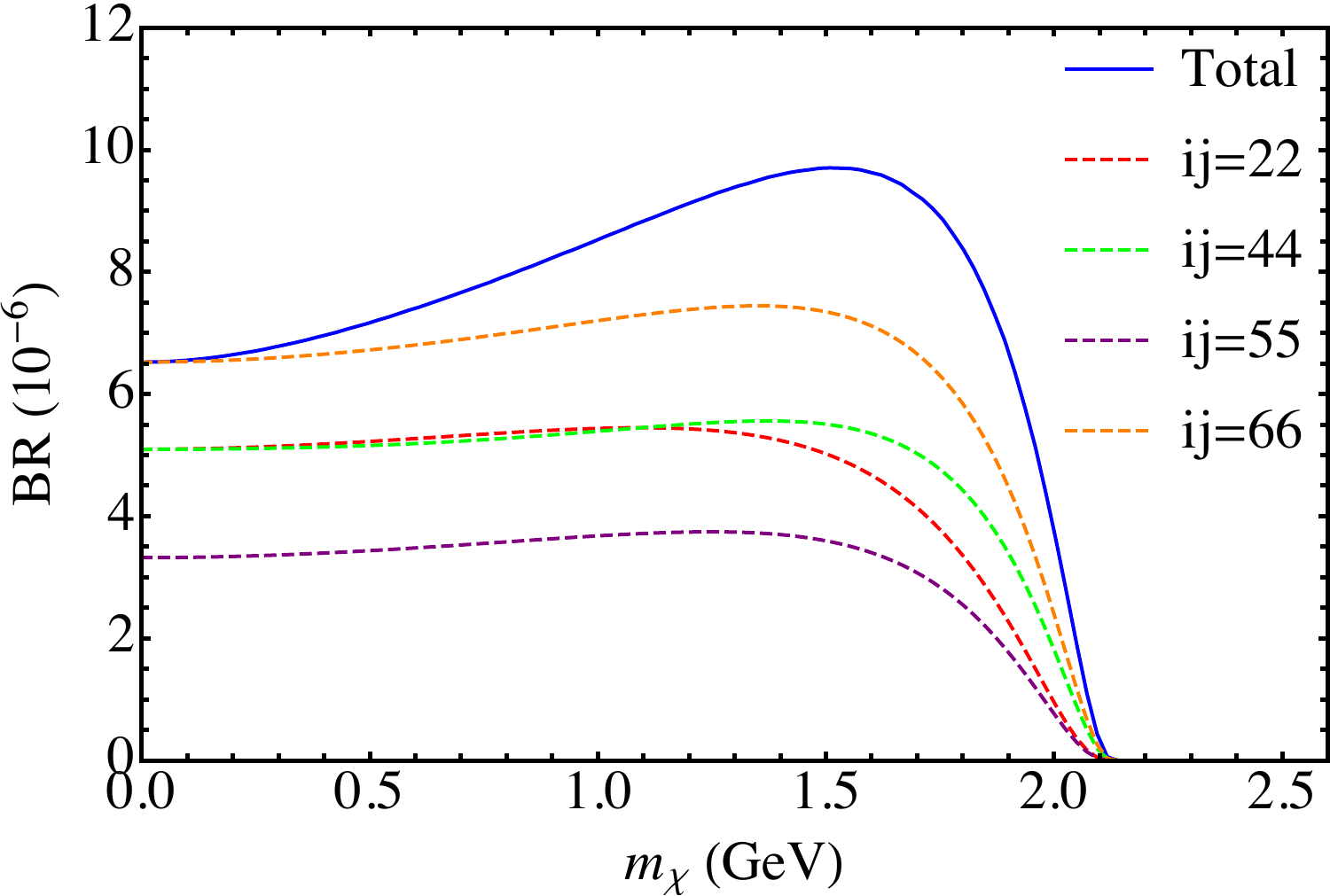}} 
	\caption{The upper limits of branching ratios of $B_c$ decays to Majorana fermions.}
	\label{br-1-7}
\end{figure}

With the effective coupling constants achieved above, we calculate the upper limits for of the branching fractions of $B_c$ decays. The results are shown in Fig. \ref{br-1-7}, where the dashed lines represent those calculated in the first way and the solid line corresponds to that of the second way. One can see that the results of two different ways do not coincide in most $m_\chi$ regions. The difference comes from the contribution of the interference terms. For the $B_c^-\to P$ processes, the three cases $ij=11$, $ij=33$, and $Total$ have the same value when $m_{\chi}=0$. At some points, $ij=55$ coincides with $Total$. For the $B_c^-\to V \chi\chi$ processes, $ij=66$ coincides with $Total$ when $m_{\chi}=0$. One notices that as $m_\chi$ increases, the branching ratios (for $Total$) first increase slowly and then decrease rapidly. This is a result of the competition between the phase space and the effective coupling constants. The upper limits of the branching fractions of $B_c\to P \chi\chi$ are 1 order of magnitude less than those of $B_c^-\to V \chi\chi$, which is mainly due to different experimental bounds in Table~\ref{tab1}. The upper limits of branching ratios will be scaled down, as more precise experimental limits are obtained in the future. 

\begin{figure}[htbp]
	\centering
	\subfigure[~$B_c^- \to D_s^- \bar\chi\chi$]{
		\label{dq2-}
		\includegraphics[width=0.45\textwidth]{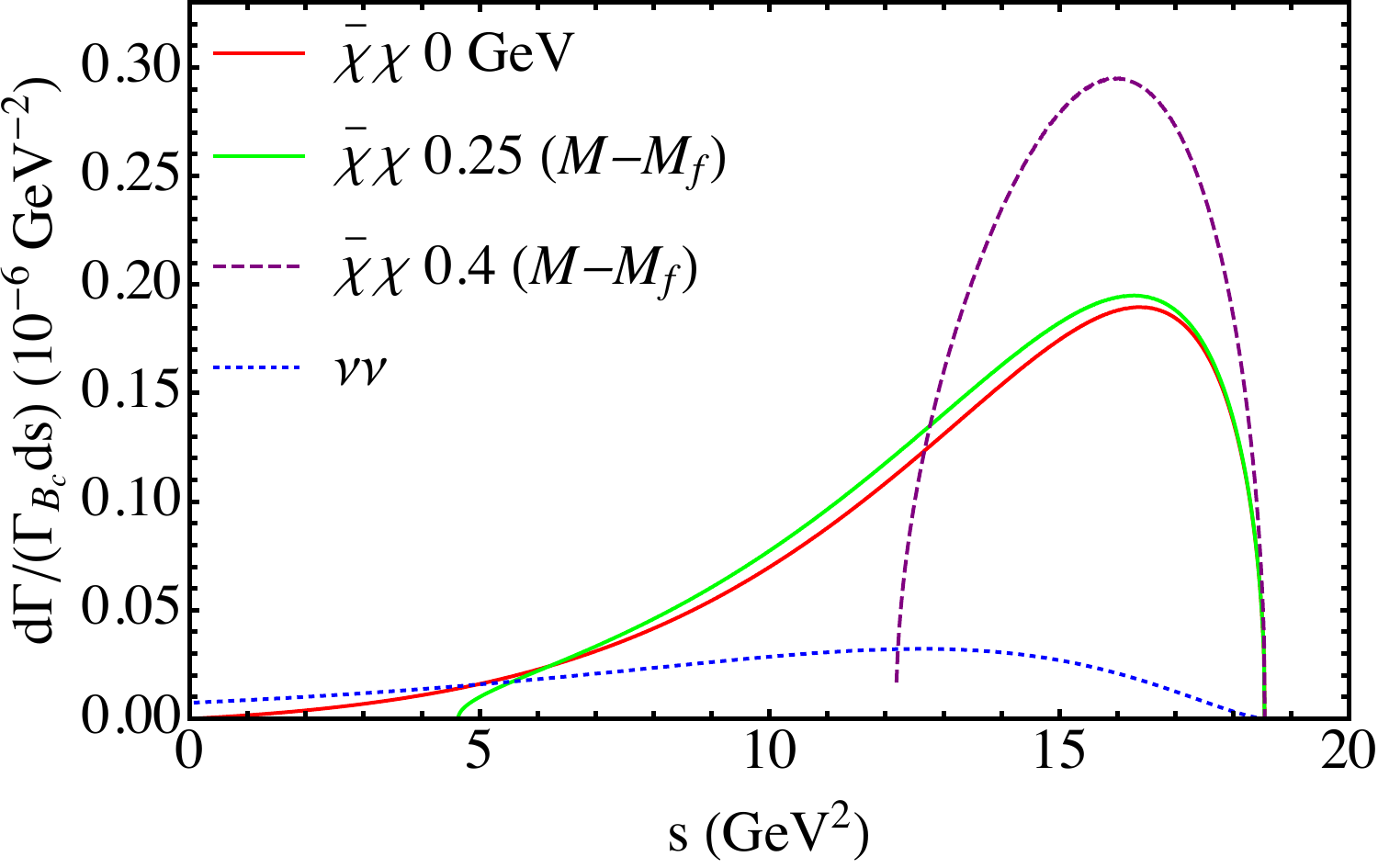}} 
	\hspace{0.5cm}
	\subfigure[~$B_c^- \to D^- \bar\chi\chi$]{
		\label{dq2-2}
		\includegraphics[width=0.45\textwidth]{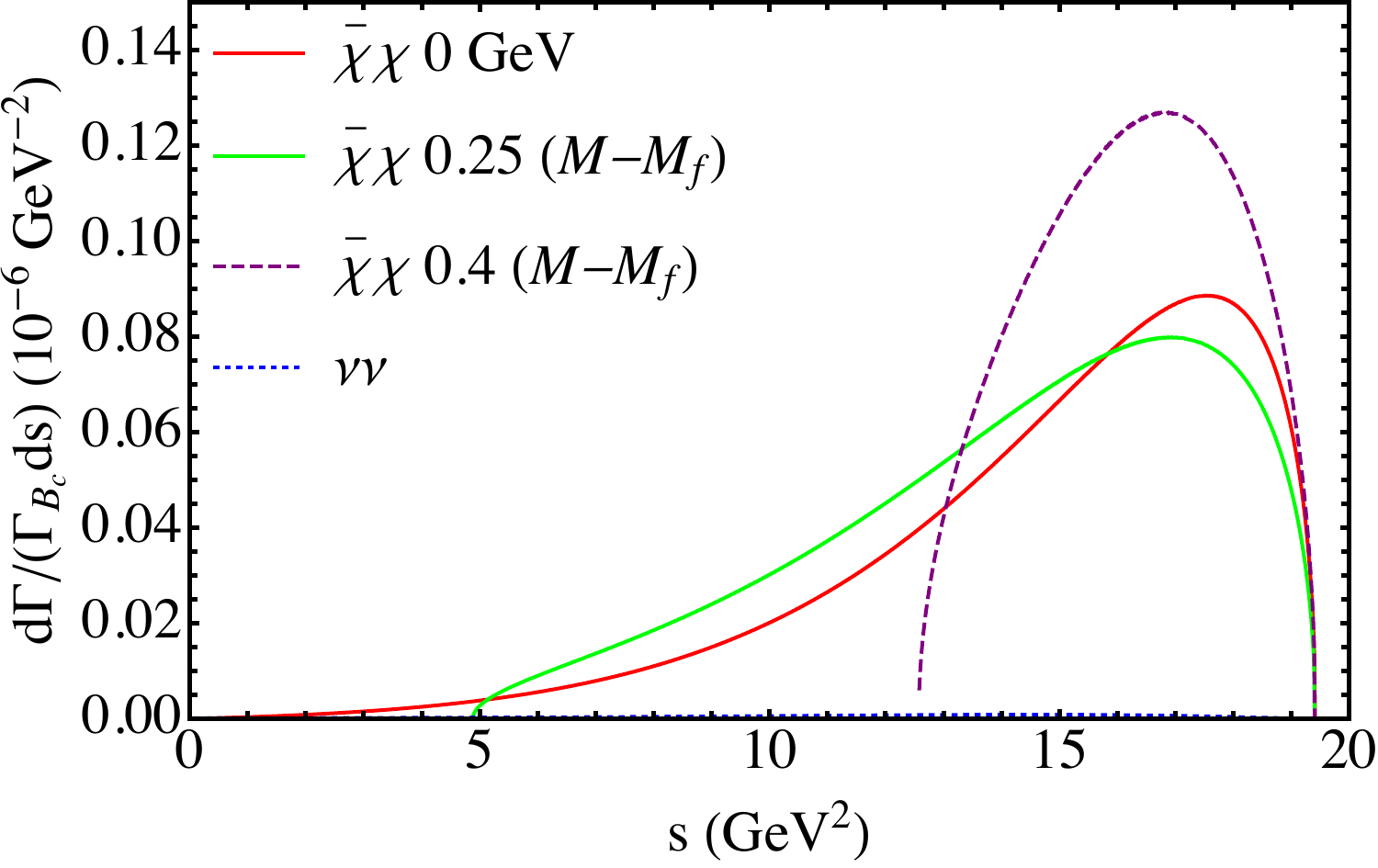}} \\
	\subfigure[~$B_c^- \to D_s^{*-} \bar\chi\chi$]{
		\label{dq2-3}
		\includegraphics[width=0.45\textwidth]{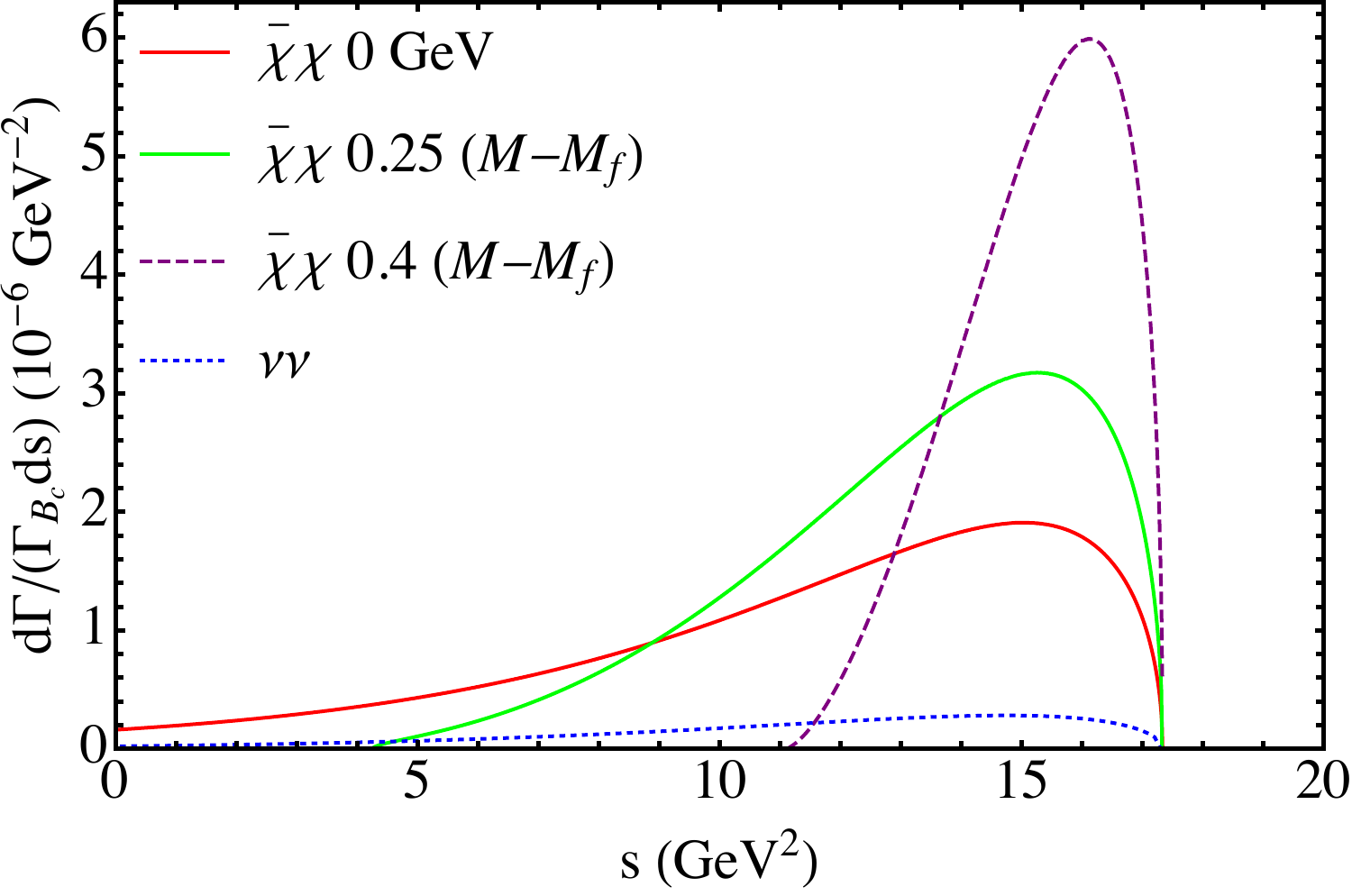}} 
	\hspace{0.5cm}
	\subfigure[~$B_c^- \to D^{*-} \bar\chi\chi$]{
		\label{dq2-5}
		\includegraphics[width=0.45\textwidth]{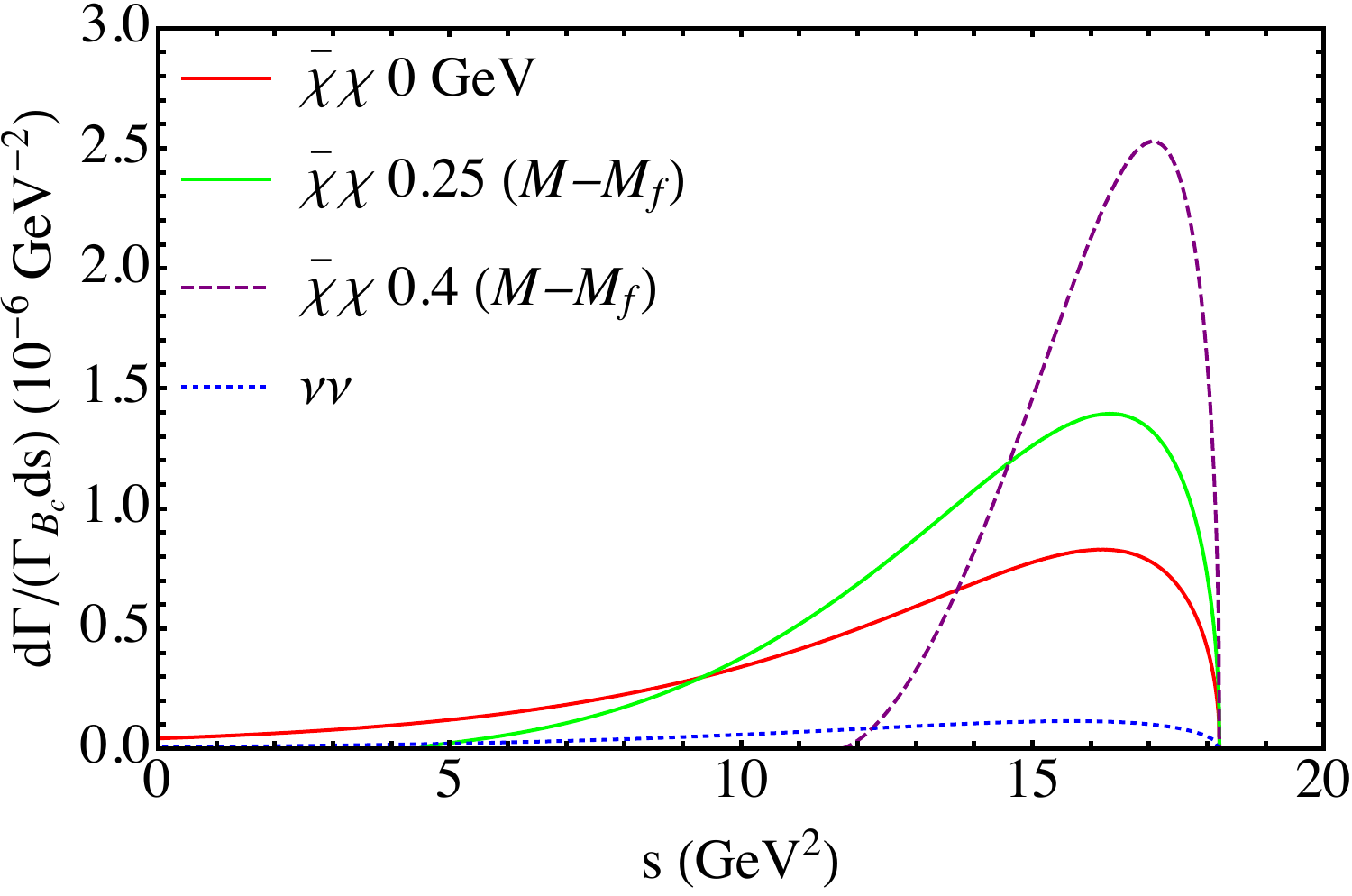}}
	\caption{The differential branching ratios of $B_c$ decays to Majorana fermions.}
	\label{dq2-1-5}
\end{figure}

In Fig.~\ref{dq2-1-5} we present the differential branching fractions as functions of $s$ which is defined as $s=(P-P_f^{})^2$. As examples, three cases with $m_\chi=0$ GeV, $0.25(M-M_f)$, and $0.4(M-M_f)$, respectively, are considered. For comparison, the SM background with $\bar\nu \nu$ emission is also plotted as blue dashed lines, which are less than those of the $\bar\chi\chi$ emission channels in most regions of $s$. The left starting point of the curves is the lower bound of $s$, which is determined by the mass of $\chi$. The position of peaks of the distribution curves is almost independent of $m_\chi$, which is at the region $s=16\sim18$ $\rm GeV^2$. We can see that the peak value gets larger as $m_\chi$ increases, because the branching ratio increases with $m_\chi$ (until reaches its maximum value around $m_\chi=1.5$ GeV) .

\subsection{$\chi$ is a Dirac fermion}

\begin{figure}[htbp]
	\centering
	\subfigure[~$B_c^- \to D_s^- \bar\chi\chi$]{
		\label{width-10}
		\includegraphics[width=0.45\textwidth]{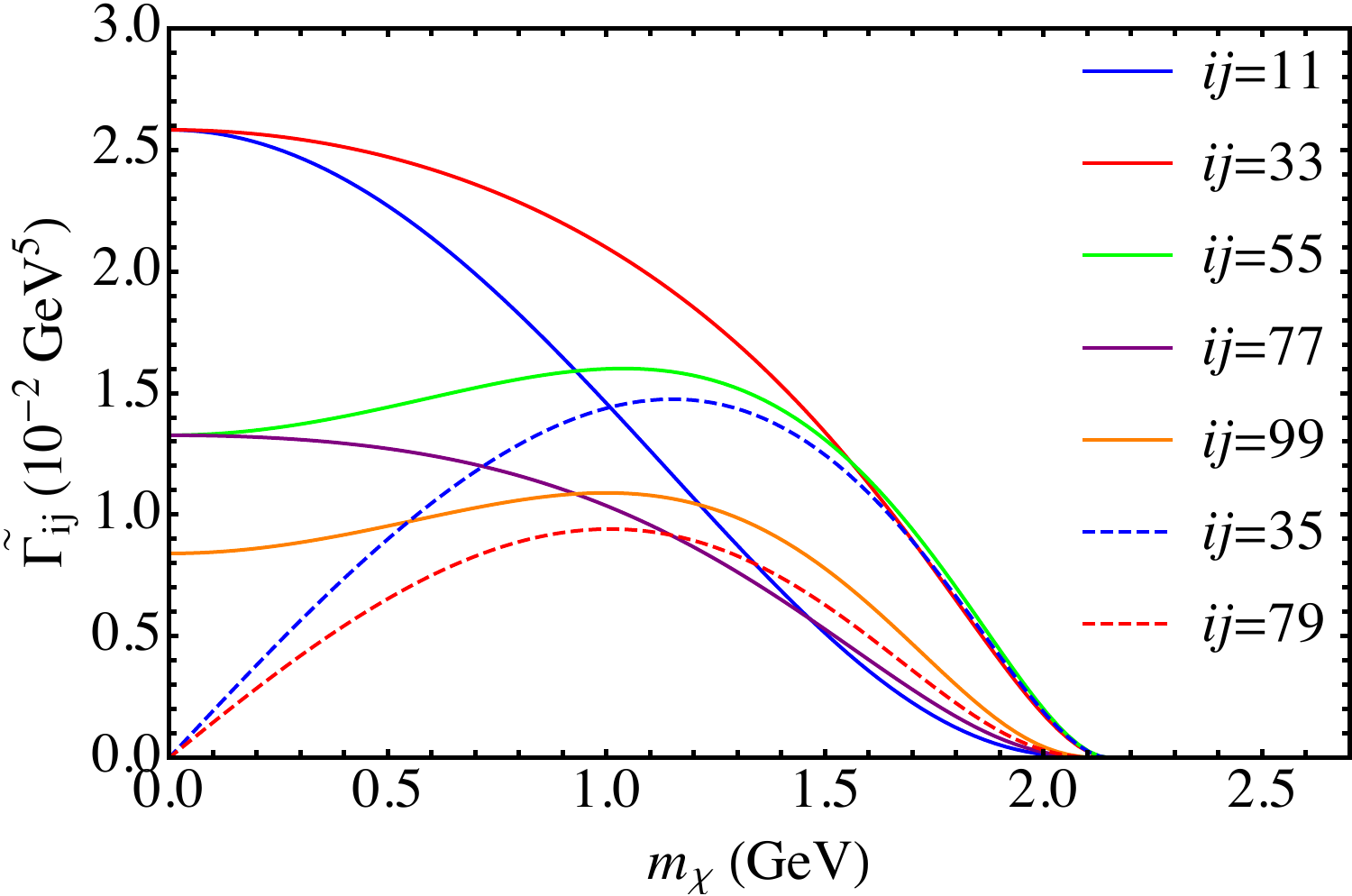}} 
	\hspace{0.5cm}
	\subfigure[~$B_c^- \to D_s^{*-} \bar\chi\chi$]{
		\label{width-16}
		\includegraphics[width=0.45\textwidth]{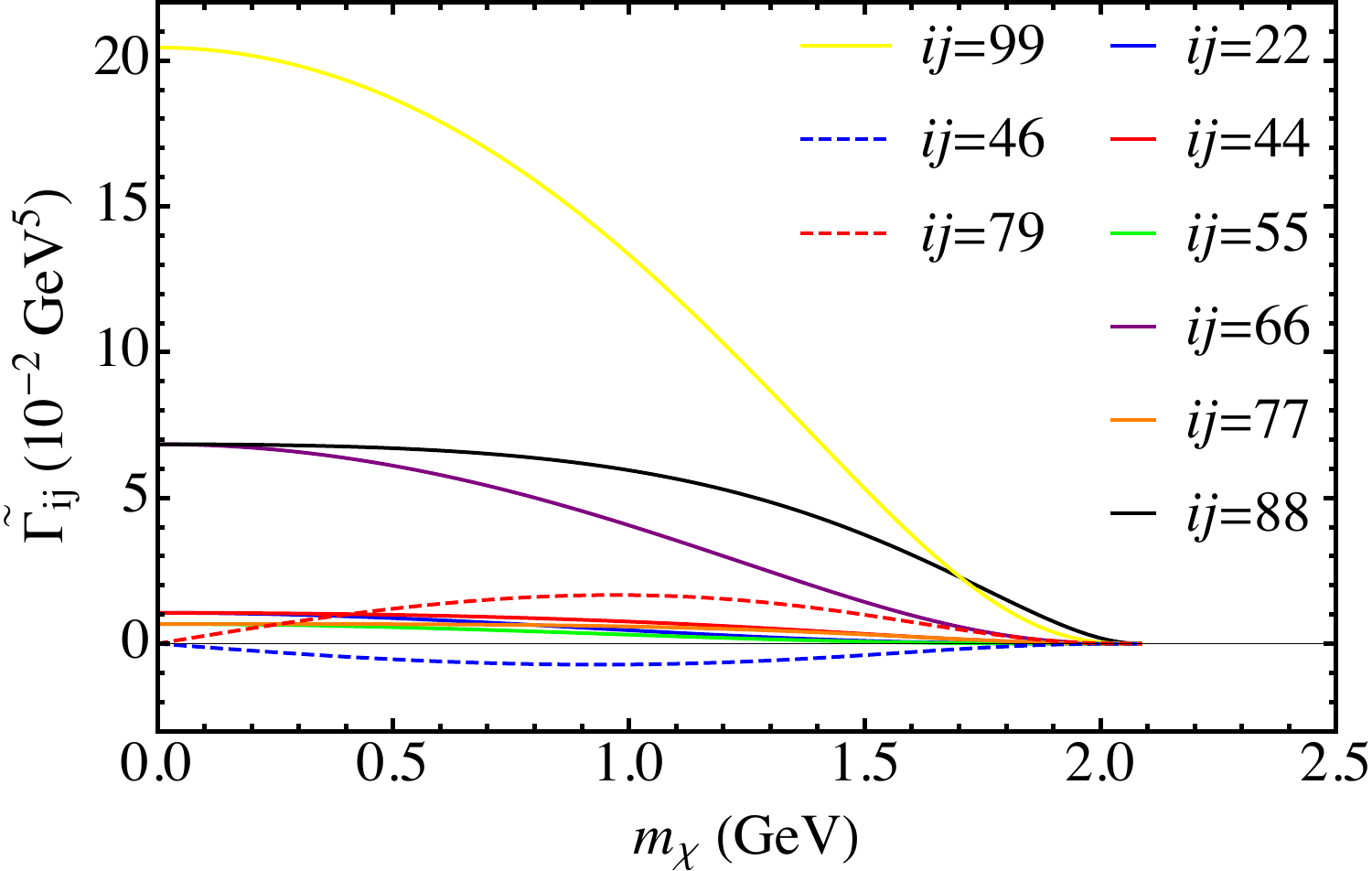}} \\
	\subfigure[~$B_c^- \to D^- \bar\chi\chi$]{
		\label{width-12}
		\includegraphics[width=0.45\textwidth]{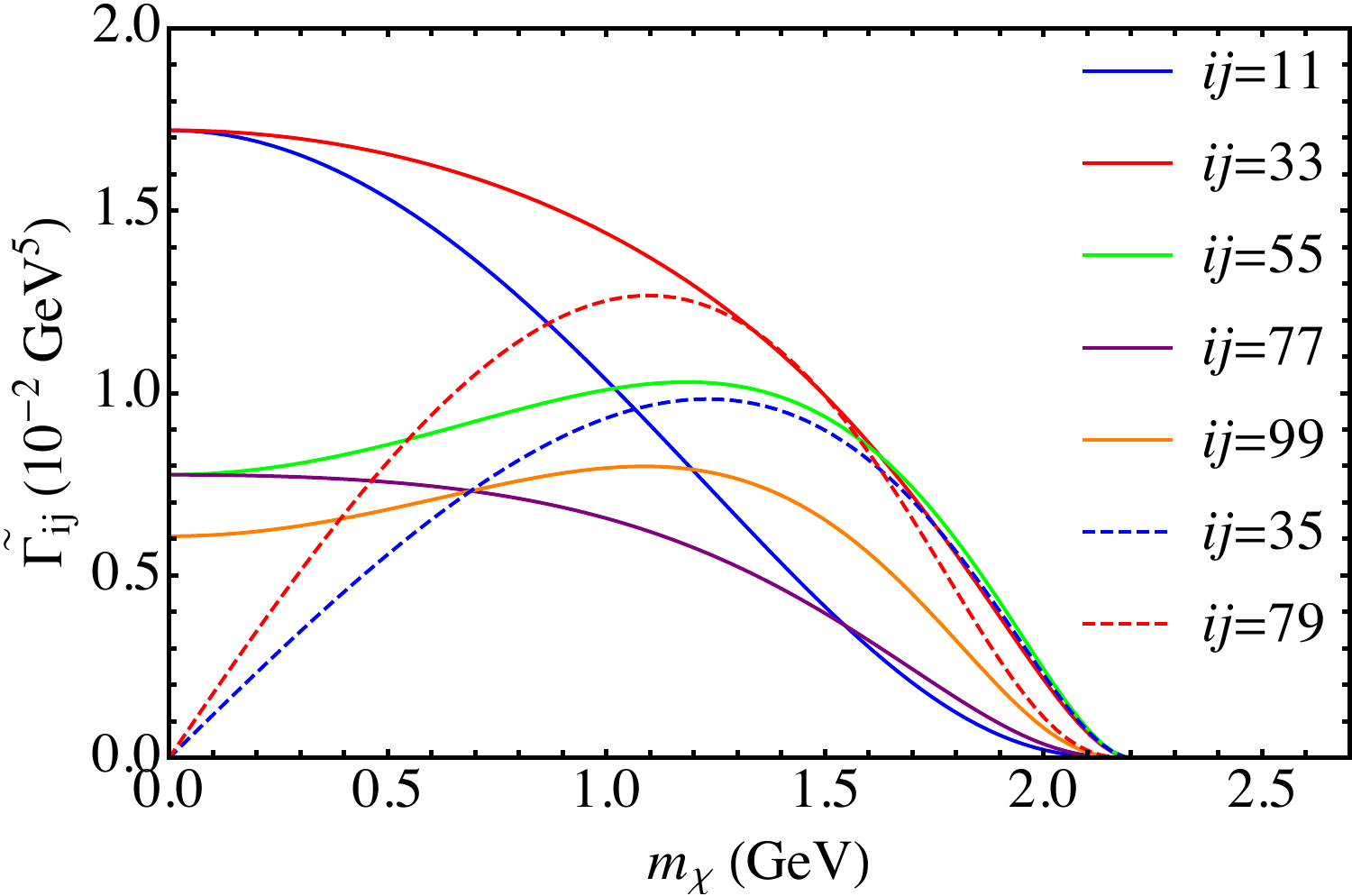}} 
	\hspace{0.5cm}
	\subfigure[~$B_c^- \to D^{*-} \bar\chi\chi$]{
		\label{width-18}
		\includegraphics[width=0.45\textwidth]{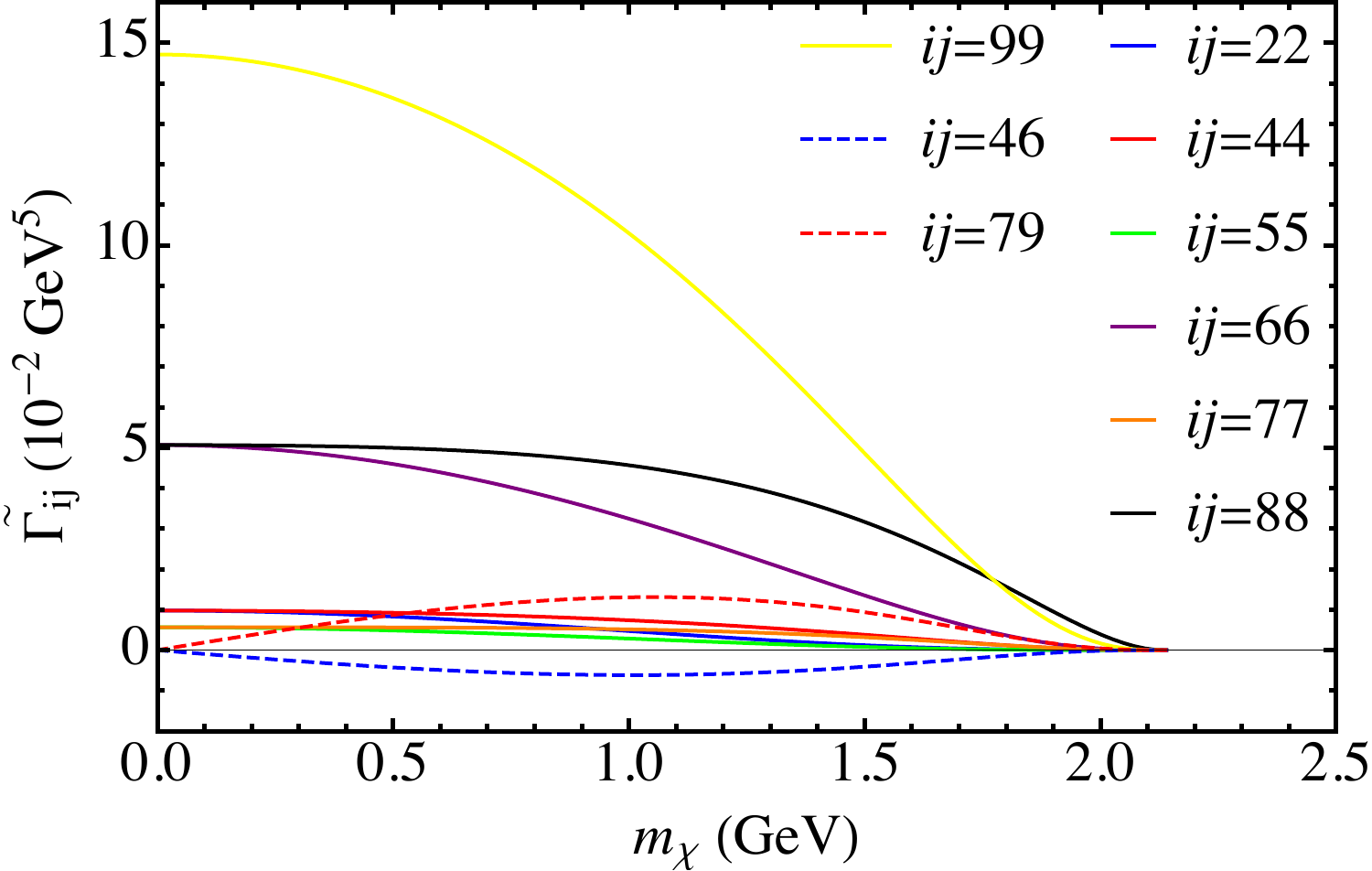}} \\
	\subfigure[~$B_c^- \to B^- \bar\chi\chi$]{
		\label{width-14}
		\includegraphics[width=0.45\textwidth]{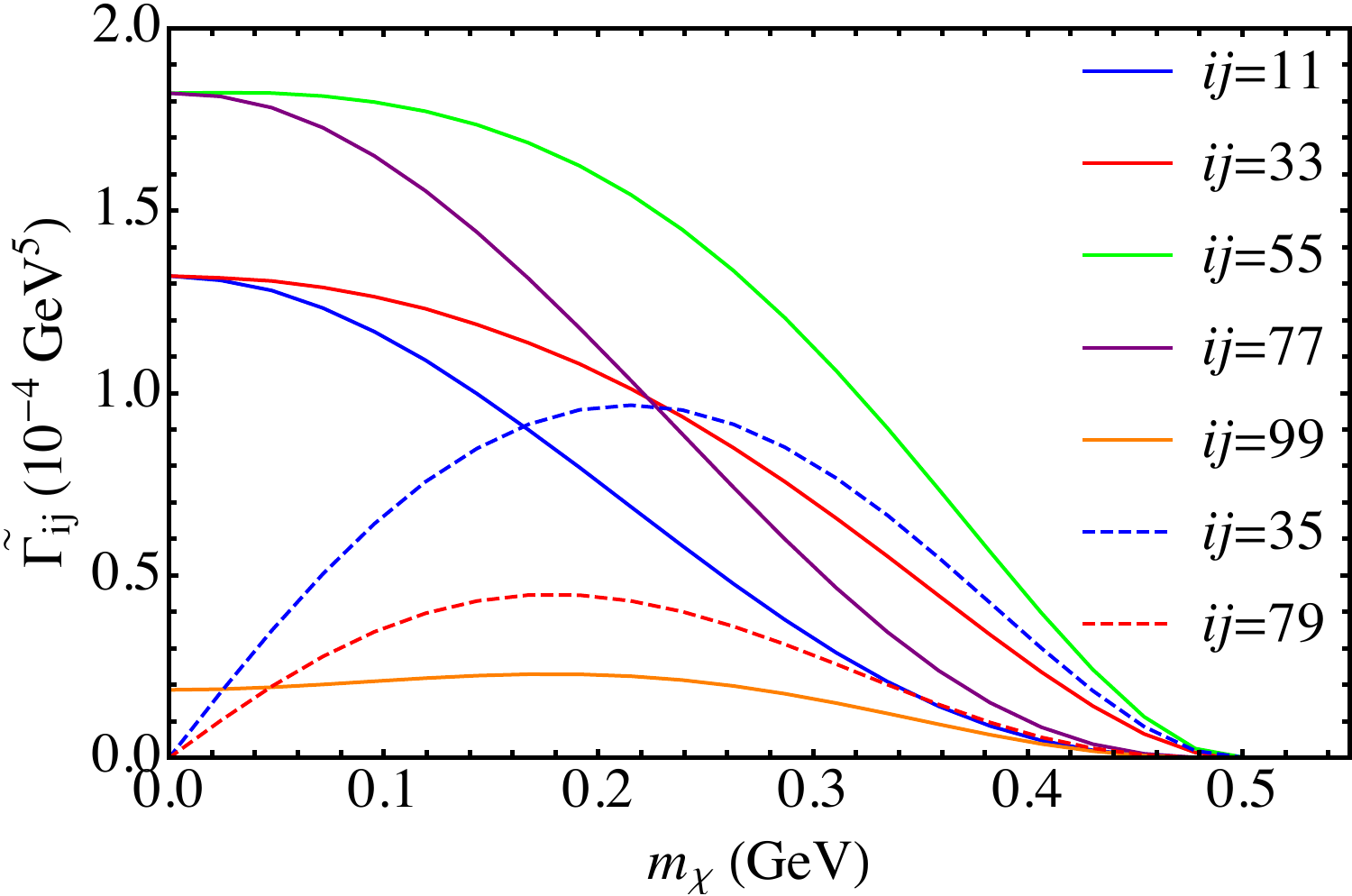}} 
	\hspace{0.5cm}
	\subfigure[~$B_c^- \to B^{*-} \bar\chi\chi$]{
		\label{width-20}
		\includegraphics[width=0.45\textwidth]{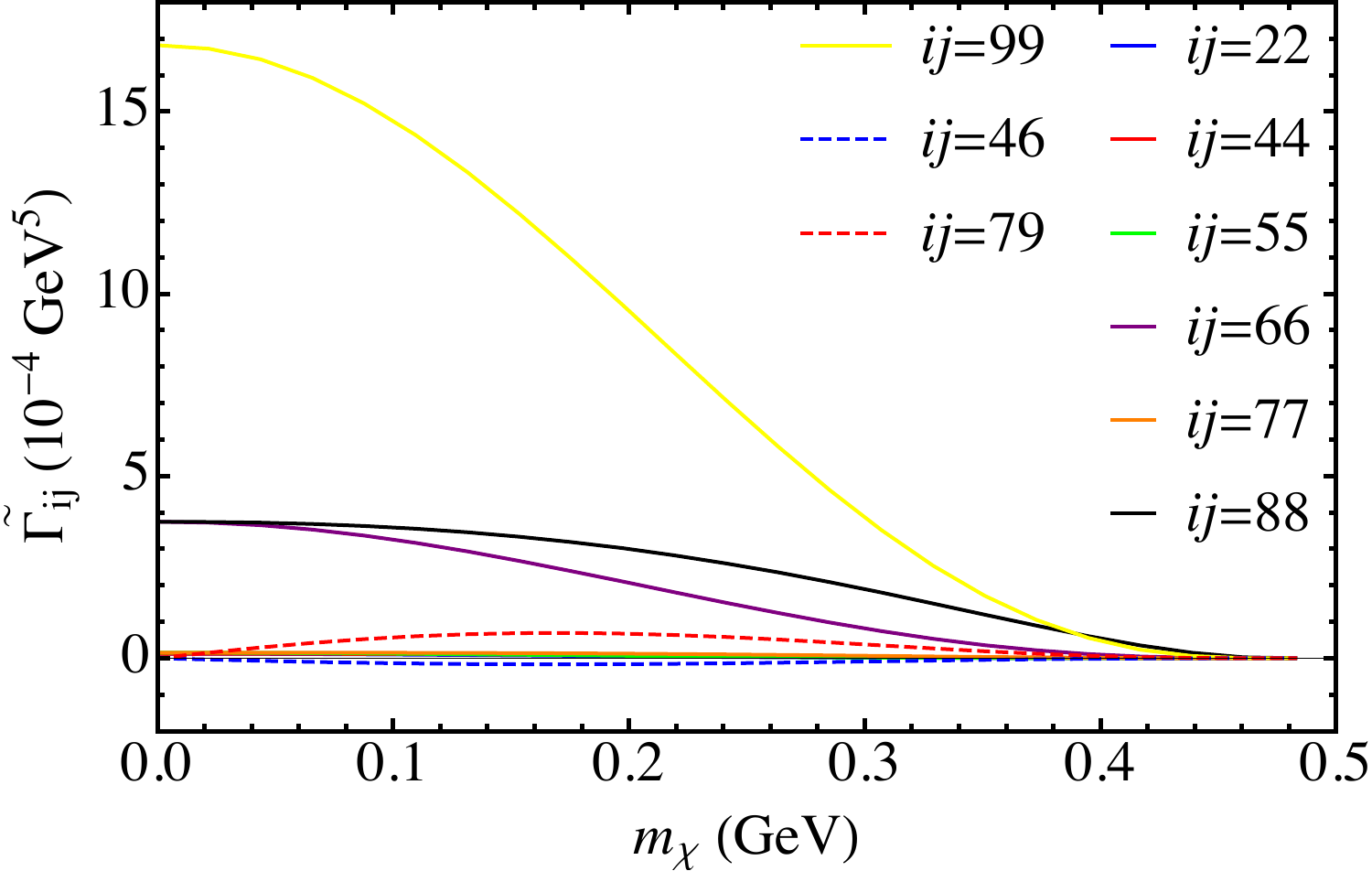}} 
	\caption{$\tilde\Gamma_{ij}$ for $B_c\rightarrow h^{(\ast)}\bar\chi\chi$ with $\chi$ being a Dirac fermion.}
	\label{width-10-20}
\end{figure}

\begin{figure}[htbp]
	\centering
	\subfigure[~$B_c^- \to D_s^{-} \bar\chi\chi$]{
		\label{br-2}
		\includegraphics[width=0.45\textwidth]{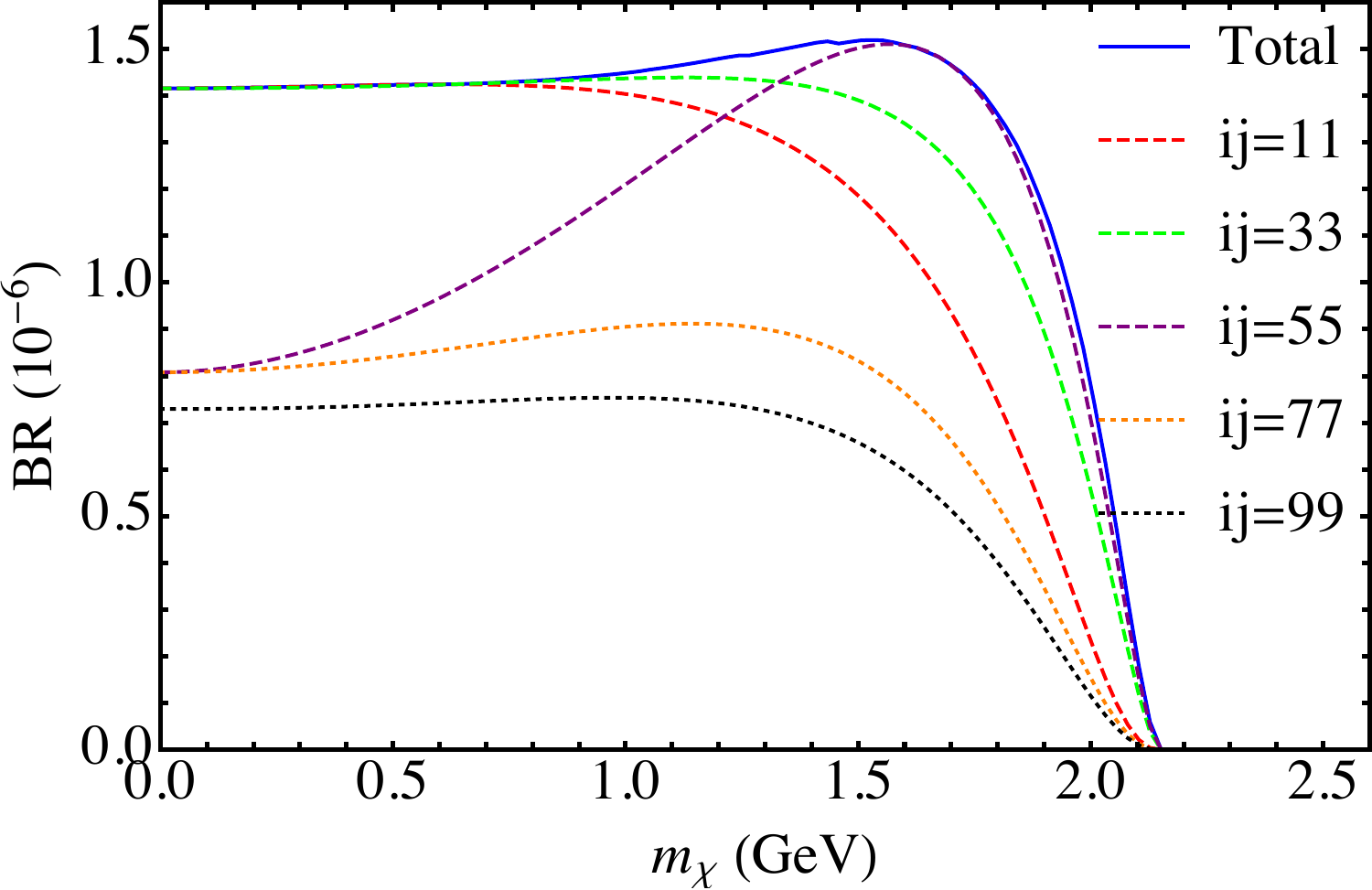}} 
	\hspace{0.5cm}
	\subfigure[~$B_c^- \to D^{-} \bar\chi\chi$]{
		\label{br-4}
		\includegraphics[width=0.45\textwidth]{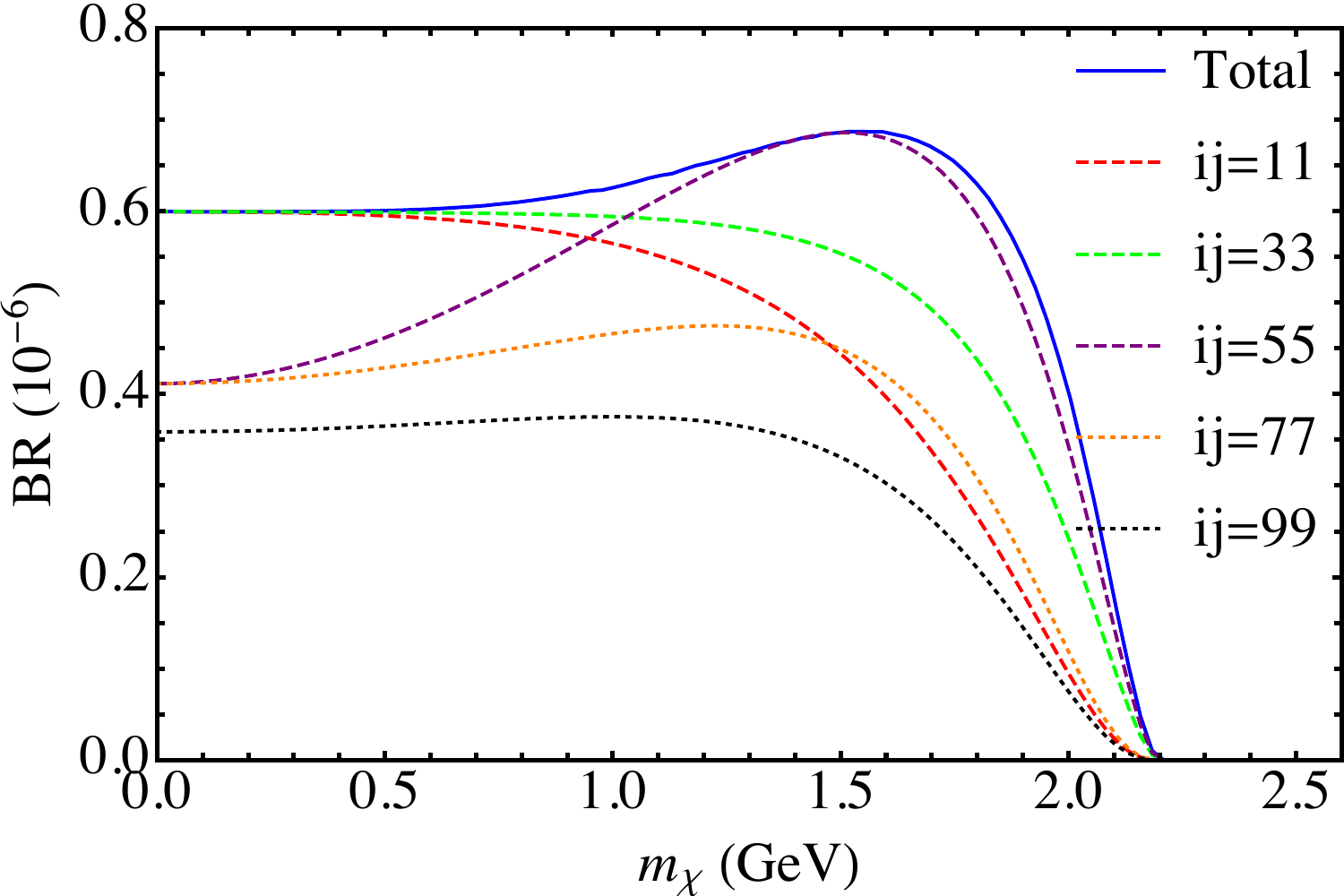}}\\
	\subfigure[~$B_c^- \to D_s^{*-} \bar\chi\chi$]{
		\label{br-6}
		\includegraphics[width=0.45\textwidth]{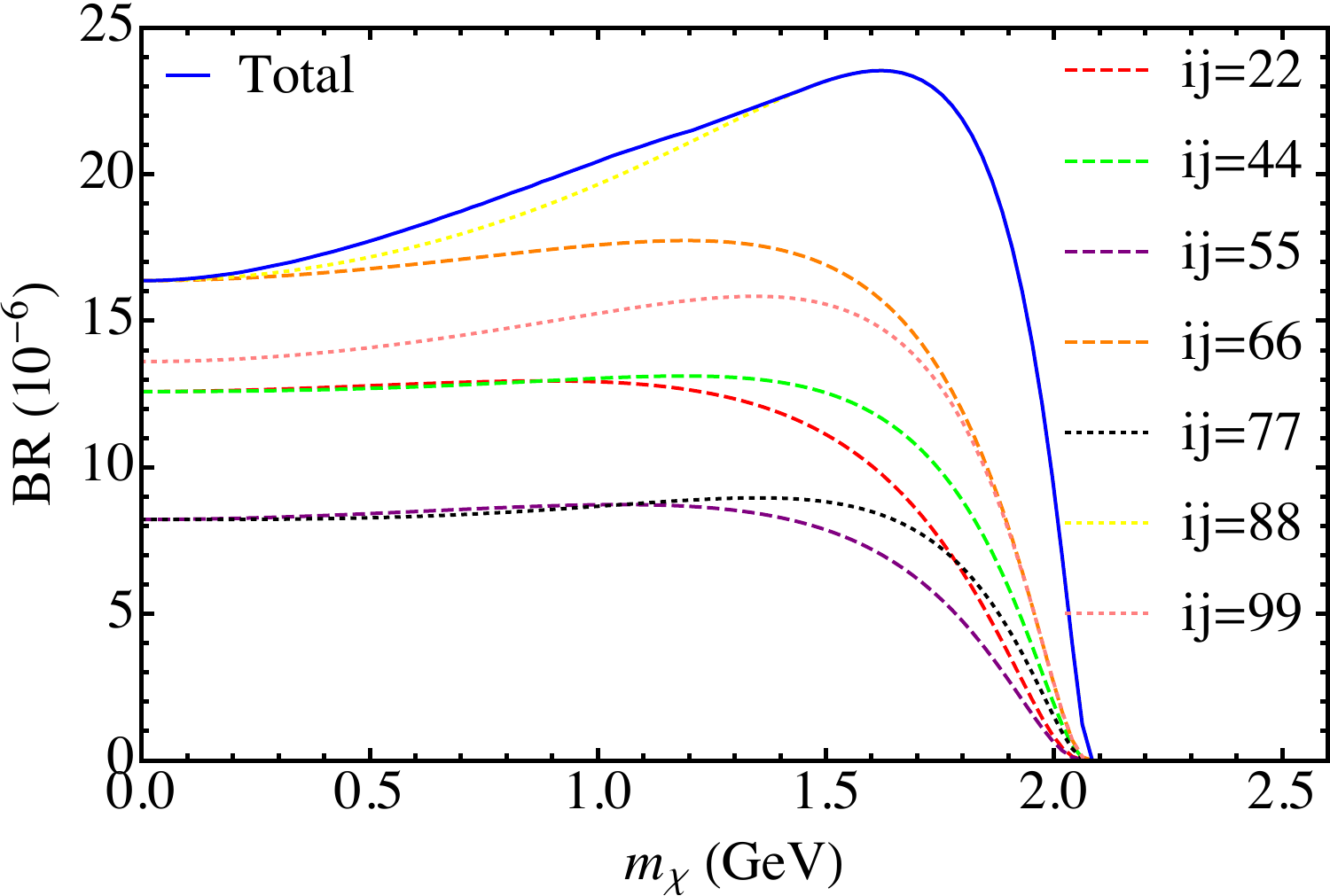}} 
	\hspace{0.5cm}
	\subfigure[~$B_c^- \to D^{*-} \bar\chi\chi$]{
		\label{br-8}
		\includegraphics[width=0.45\textwidth]{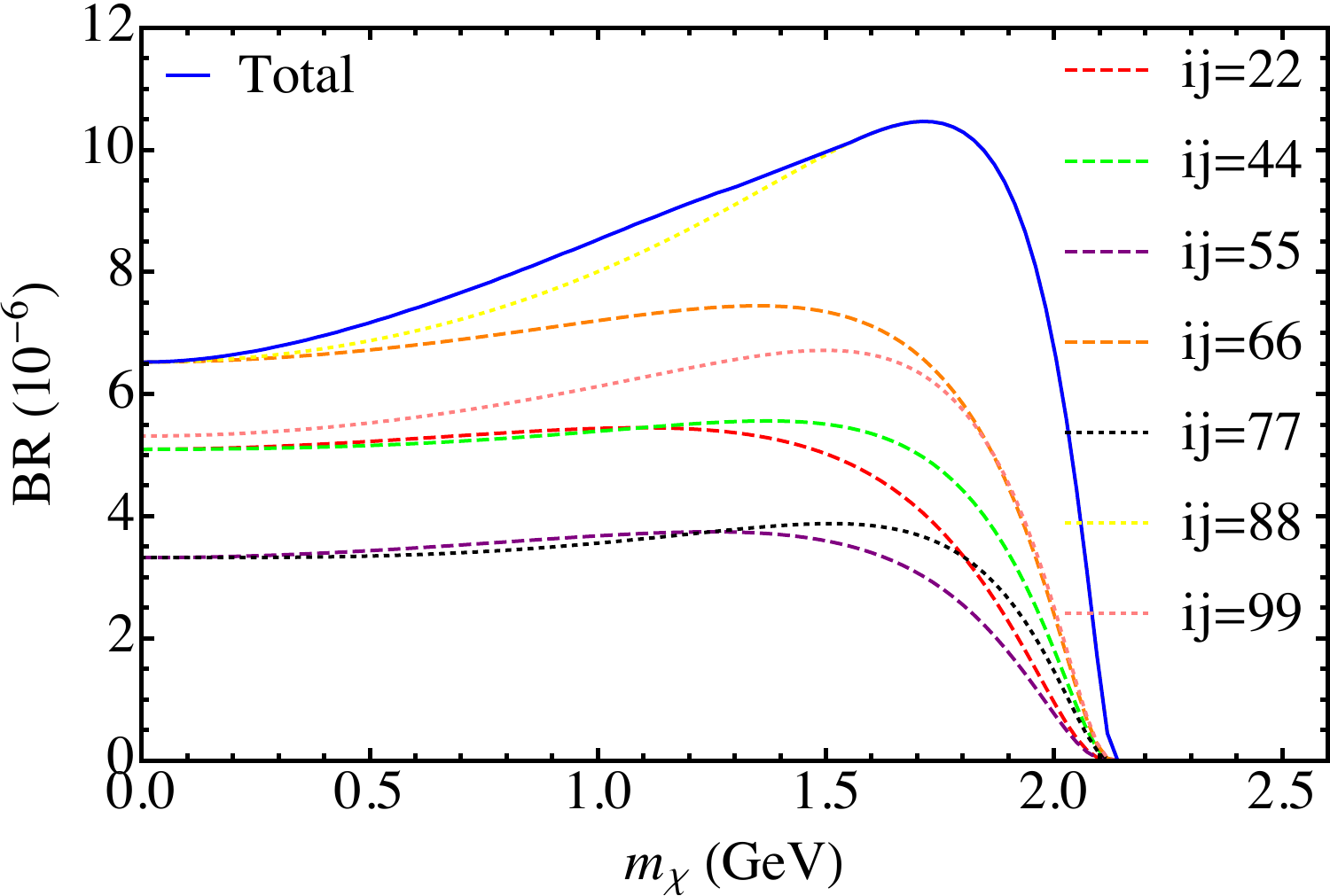}}
	\caption{The upper limits of branching ratios of $B_c$ decays to Dirac fermions.}
	\label{br-2-8}
\end{figure}

The similar analysis can also be applied to the Dirac fermions. The effective Lagrangians take the same forms as those in Eqs.~(\ref{eq15}) and (\ref{eq17}). In Fig.~\ref{width-10-20}, we plot $\widetilde\Gamma_{ij}$ as the function of $m_\chi$, which is about half of the corresponding one  in the Majorana case. One can see there are several additional terms $\widetilde\Gamma_{77}$, $\widetilde\Gamma_{88}$, and $\widetilde\Gamma_{99}$ which do not exist in the Majorana case. The effective coupling constants obtained by comparing with the experimental results are used to find the maximum values of the branching fractions which are plotted in Fig.~\ref{br-2-8}. In Figs.~\ref{br-2} and \ref{br-4}, we give the results of the ${B_c}\to P\bar\chi\chi$ processes. If we only consider the contribution of $O_7$ or $O_9$, one can see the upper bound of the branching ratios, which are labeled by $ij=77$ and $ij=99$, respectively, are less than those resulted by other operators. This means that they do not affect the maximum branch fractions obtained by the second way, namely considering the operators altogether. This leads to the result that the upper limits of the branching ratios of such channels are the same as those in the Majorana case. Correspondingly, the differential branching fractions of two cases are also the same with each other. 

In Figs.~\ref{br-6} and \ref{br-8},  the results of the ${B_c}\to V\bar\chi\chi$ processes are presented. The upper bound labeled by $Total$ (solid blue line) is the same as that in Fig.~\ref{br-5} or \ref{br-7} when $m_{\chi}$ is less than $1.2$ or $1.3$ GeV. It becomes larger when $m_{\chi}$ continues to increase, as the $\widetilde\Gamma_{77}$ term, which does not exist in the Majorana case, will give the main contribution. So, in this range, there are some differences between the upper bounds obtained in two cases (of course, if only $O_7$, $O_8$, and $O_9$ give contribution, the Majorana case is not allowed). The errors of $\widetilde\Gamma_{ij}$ are about $\pm 20\% $ from LCSR with uncertainties of parameters in Table~III. It will affect the upper limits of the coupling constants. By varying the parameters in BS method by $\pm 5\%$, the errors are from $\pm 7\% $ to $\pm 17\%$. Total errors of branching ratios of $B_c$ meson decays in Figs.~\ref{br-1-7} and \ref{br-2-8} are about $\pm 30\% $.

\begin{figure}[htbp]
	\centering
	\subfigure[~$B_c^- \to D_s^{*-} \bar\chi\chi$]{
		\label{dq2-4}
		\includegraphics[width=0.44\textwidth]{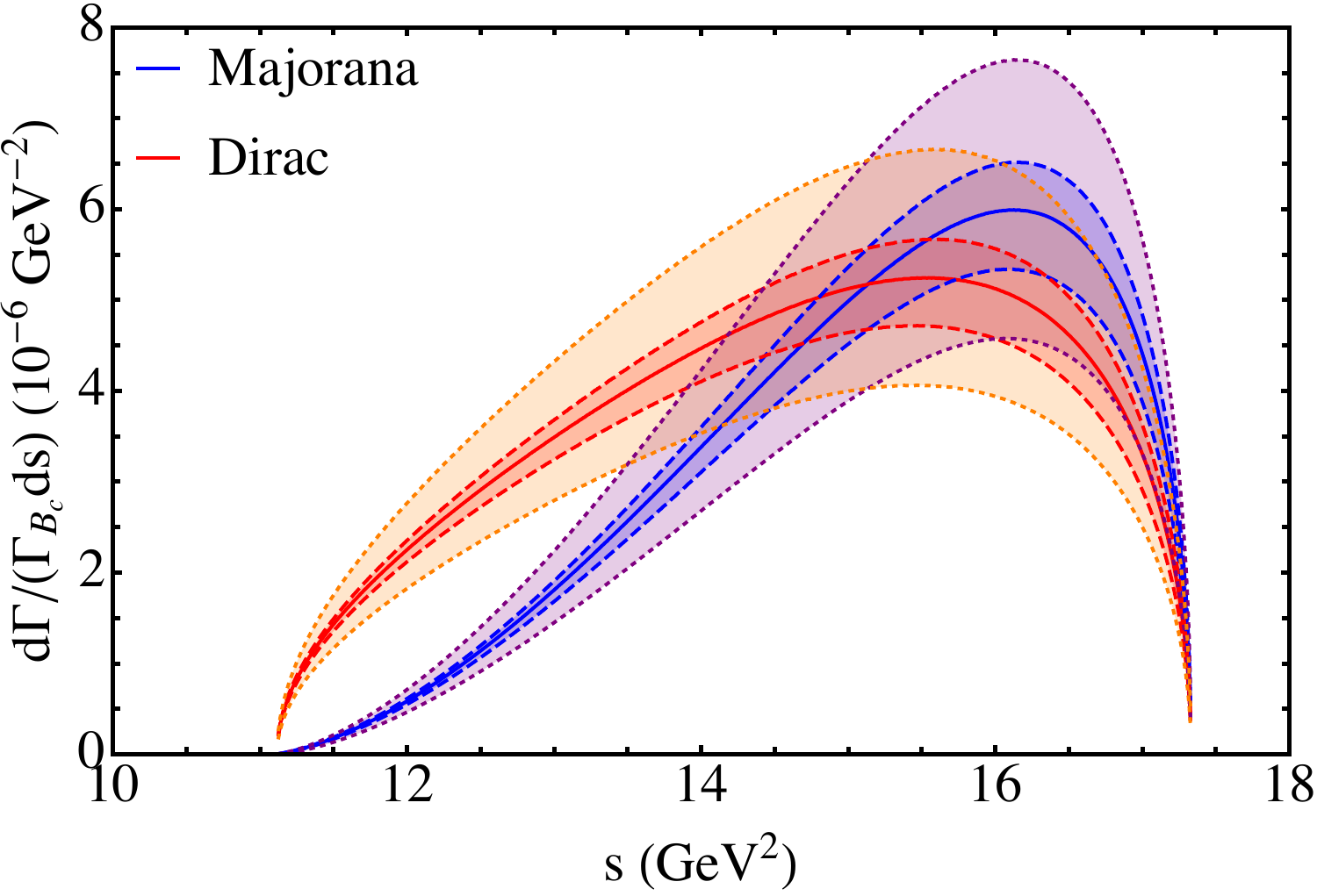}} 
	\hspace{0.5cm}
	\subfigure[~$B_c^- \to D^{*-} \bar\chi\chi$]{
		\label{dq2-6}
		\includegraphics[width=0.45\textwidth]{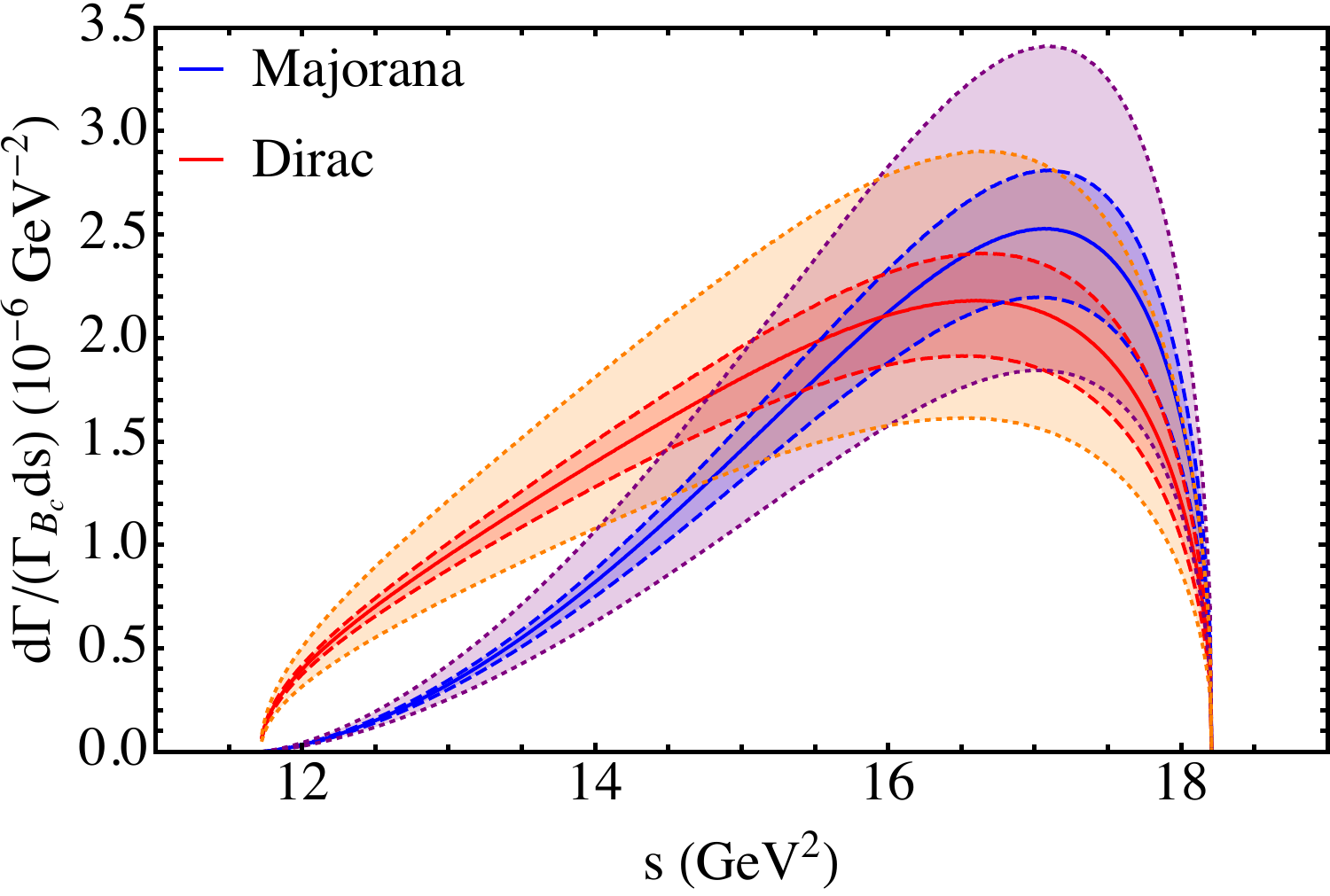}}
	\caption{The differential branching ratios of $B_c$ decays with $m_\chi=0.4~(M-M_f)$. The dashed lines represent the errors come from BS method. The dotted lines represent the errors come from both BS and LCSR. }
	\label{dq2-4-6}
\end{figure}

Correspondingly, the differential branching ratios, which are plotted in Fig.~\ref{dq2-4-6}, should also show some differences with those in the Majorana case. When $m_\chi=0.4(M-M_f)$, the distribution curves have clearly different shapes from those in Figs.~\ref{dq2-3} and \ref{dq2-5}. By varying the parameters in BS method by $\pm 5\%$, the errors of distribution curves are less than $\pm 10 \%$. Further considering the uncertainties from LCSR in Table~\ref{tab3}, the total errors are about $\pm 30\%$. The distributions of Majorana and Dirac types still can be distinguished in regions of $11-13.5~\rm{GeV}^2$ ($B_c\to D_s^*$) or $12-14~\rm{GeV}^2$ ($B_c\to D^*$). This might provide a way to distinguish between them. It is necessary to notice that these results are upper limits when we assume that all operators contribute at the same time. If only a few operators contribute to this process, for example, $Q_7$, $Q_8$, and $Q_9$, the distinction between Majorana and Dirac will become very obvious. In such cases, only the Dirac type final fermions are allowed. 

\section{Conclusion}

We studied the FCNC processes of the $B$ meson decaying to the invisible spin-1/2 fermions. Both the Majorana and Dirac cases were considered. The effective Lagrangians were introduced to describe the coupling between invisible particles and quarks. By comparing the theoretical predictions of BR$(B\rightarrow D_{(s)}^{(\ast)}\bar\nu\nu)$ and the experimental upper bounds for BR$(B\rightarrow D_{(s)}^{(\ast)}\slashed E)$, we derived the constraints of the effective coupling constants. By scanning the allowed parameter space, we derived the upper limits of the branching fractions for the similar processes of $B_c$ meson. When the final meson was a pseudoscalar, the upper limits of the branching fractions was of the order of $10^{-6}$, and when the final meson was vector, it was of the order of $10^{-5}$. These results were much larger than those of the SM background. The differential branching fractions of Majorana and Dirac invisible particles were of different shapes when $m_\chi$ was larger than $1.2$ or $1.3$ GeV. This could provide a way to distinguish between the Majorana type particle from the Dirac one.

\section{Acknowledgments}

This work was supported in part by the National Natural Science Foundation of China under Grant No.~11575048. We also thank the HPC Studio at Physics Department of Harbin Institute of Technology for access to computing resources through INSPUR-HPC@PHY.HIT.

\bibliography{reference}

\end{document}